\documentclass{aa}
\usepackage{graphicx}
\usepackage[varg]{txfonts}
\usepackage{physics}
\usepackage{amsmath, amssymb}
\usepackage{dsfont}
\usepackage{bbm}
\usepackage{xcolor}
\usepackage{ulem}
\usepackage{xpatch}
\usepackage{hyperref}
\usepackage{import}
\usepackage{makeidx}
\usepackage{changes}
\usepackage{hhline}

\newcommand{\colout}[1]{\bgroup\markoverwith{\textcolor{#1}{\rule[.5ex]{2pt}{0.4pt}}}\ULon}

\makeatletter
\renewcommand*\aa@pageof{, page \thepage{} of \pageref*{LastPage}}
\newcommand{\pder}[2][]{\frac{\partial#1}{\partial#2}}
\makeatother

\renewcommand{\arraystretch}{1.5}

\makeindex

\begin{document}


%
\titlerunning{Inner dead zone edge variability}
\authorrunning{M. Cecil and M. Flock}
\title{Variability of the inner dead zone edge in 2D radiation hydrodynamic simulations} 
%
%
\author{
 Michael~Cecil~and~Mario~Flock
}
\institute{
 Max-Planck Institute for Astronomy (MPIA), Königstuhl 17, 69117 Heidelberg, Germany \\
 e-mail: \texttt{cecil@mpia.de}
%
}
\date{Received ....; accepted ....}

\abstract
{The inner regions of protoplanetary disks are prone to thermal instability (TI), which can significantly impact the thermal and dynamical evolution of planet-forming regions. Observable as episodic accretion outbursts, such periodic disturbances shape the disk's vertical and radial structure.}
{We investigate the stability of the inner disk edge around a Class II T Tauri star and analyse the consequences of TI on the thermal and dynamic evolution in both the vertical and radial dimensions. A particular focus is laid on the emergence and destruction of solid-trapping pressure maxima. }
{We conduct 2D axisymmetric radiation hydrodynamic simulations of the inner disk in a radial range of 0.05 AU to 10 AU. The models include a highly turbulent inner region, the transition to the dead zone, heating by both stellar irradiation and viscous dissipation, vertical and radial radiative transport and tracking of the dust-to-gas mass ratio at every location. The simulated time frames include both the TI phase and the quiescent phase between TI cycles. We track the TI on S-curves of thermal stability.}
{TI can develop in disks with accretion rates of $\geq 3.6\cdot 10^{-9}~\mathrm{M}_\odot~\mathrm{yr}^{-1}$ and results from the activation of the magnetorotational instability (MRI) in the dead zone after the accumulation of material beyond the MRI transition. The TI creates an extensive MRI active region around the midplane and disrupts the stable pebble- and migration trap at the inner edge of the dead zone. Our simulations consistently show the occurrence of TI-reflares, which, together with the initial TI, produce pressure maxima in the inner disk within 1 AU, possibly providing favourable conditions for streaming instability. During the TI phase, the dust content in the ignited regions adapts itself in order to create a new thermal equilibrium manifested on the upper branch of the S-curve. In these instances, we find a simple relation between the gas- and dust-surface density.}
{On a timescale of a few thousand years, TI regularly disrupts the radial and vertical structure of the disk within 1 AU. While several pressure maxima are created, stable migration traps are destroyed and reinstated after the TI phase. Our models provide a foundation for more detailed investigations into phenomena such as short-term variability of accretion rates.}

\keywords{protoplanetary disks --
                accretion, accretion disks --
                stars: protostars -- radiative transfer -- hydrodynamics
               }

\maketitle
\section{Introduction} \label{sec:introduction}
With the rising number of Earth-sized and super-Earth planets detected in close proximity to their star, with orbital periods of around ten days \citep{Mulders2015, Petigura2018, Mulders2018}, the questions about the conditions in the inner protoplanetary disk gain increasing relevance. Generally, the two pathways towards the existence of planets at such distances are the in-situ formation via trapping and growing of inwards-drifting pebbles \citep{Chatterjee2014, Jang2022} and the halting of planetary migration by a diminishment of the torques acting between the planet and the disk \citep{Faure2016, Flock2019, Chrenko2022}. For both of these processes, pressure maxima in the inner disk, capable of effectively trapping solid bodies, are required. A promising candidate for a mechanism creating such bumps in the pressure profile within $\sim$1 AU of the host star is the transition between a hot inner disk in which temperatures and ionisation levels are sufficiently high to allow for the emergence of the magnetorotational instability \citep[MRI, ][]{Balbus1991} and a cold outer disk where the levels of turbulence are significantly smaller \citep[e.g., ][]{Dzyurkevich2013, Cui2022}{}. In the context of two-dimensional axisymmetric simulations, \cite{Flock2016} and \cite{Flock2019} showed that this transition lies at distances between 0.1 and 1 AU, depending on the stellar luminosity, and indeed provides favourable conditions for halting both the pebble drift and the migration of planets. However, their models represented the radiation-hydrostatic structure of the inner disk without consideration of dynamic effects and heating by viscous dissipation. \\
Several studies indicate that the structure of the inner disk is not stable over long periods of time but undergoes phases of variation induced by a thermal instability (TI) in the inner regions of the dead zone \citep{Lin1985, Kley1999, Wunsch2005, Zhu2010a, Martin2011, Faure2014}. These regions are considered thermally unstable if a small temperature perturbation causes a significant increase in the heating rate compared to the cooling rate, leading to runaway heating \citep[][]{Armitage2019}. There are two scenarios of interest where this condition can be met in the inner regions of protoplanetary disks. The first scenario occurs when gas opacity increases sharply with temperature due to the ionization of hydrogen at temperatures of $\sim$10\textsuperscript{4} K, reducing the cooling rate and trapping heat near the midplane, \citep[commonly referred to as `classical' TI, ][]{Bell1993, Zhu2009}{}{}. In the second scenario, viscosity increases significantly with temperature due to the activation of the MRI at temperatures around 1000 K, greatly enhancing the heating rate \citep[][]{Desch2015, Zhu2009b, Steiner2021}{}{}. If either or both of these scenarios occur, the corresponding region of the disk is considered thermally unstable \citep[][]{Armitage2019}.  \\
Such instabilities are also indicated in investigations of the thermal structure of the inner disk, which often result in multiple solutions for the vertical thermal balance when the influence of viscous dissipation and the strong temperature dependence of the gas opacity are considered \citep[][]{Jankovic2021, Pavlyuchenkov2023}{}{}. This results in the typical `S-curves' of thermal stability in the $\Sigma - T_\mathrm{eff}$ plane, where $\Sigma$ is the surface density and  $T_\mathrm{eff}$ is the effective temperature at a fixed radius \citep[e.g., ][]{Martin2011, Nayakshin2024}{}{}. In this context, the TI constitutes the switch of the respective disk region from the stable lower- to the stable upper branch of the S-curve. Such a description is analogous to the effects of the well-known disk instability model for cataclysmic variables such as dwarf novae or X-ray transients \citep[DIM, see, e.g. review by ][]{Hameury2020}{}{}. The TI is induced as soon as the surface density reaches a critical value, after which a heating front is launched into the dead zone of the disk in a `snowplough'-fashion \citep[][]{Lin1985}{}{}. The sudden increase of turbulence in the massive dead zone results in a strong elevation of the accretion rate onto the central star. This is why TI of the inner disk is one of the candidates for physical processes that may be able to explain observed episodic accretion bursts in young stellar objects such as FU Ori \citep[][]{Herbig1977, Kley1999, Audard2014}{}{}.  \\
While a few 2D axisymmetric models of the TI of the inner disk exist \citep[][]{Kley1999, Zhu2009}{}{}, most theoretical studies of TI-induced episodic accretion events do not resolve the vertical structure of the inner disk \citep[e.g., ][]{Zhu2010a, Bae2013, Bae2014, Macfarlane2019, Kadam2022, Cleaver2023}{}{}. Moreover, these investigations were tailored to explain the timescale and magnitudes of massive accretion outbursts of Class 0/I young stellar objects. In such young massive disks, the activation of the TI mainly relies on the inward transport of material by gravitational instability (GI) \citep[e.g., ][]{Vorobyov2006}{}{}. However, both theoretical models and observations have shown that less massive Class II objects are prone to TI as well \citep[][]{Kospal2016, Steiner2021, Chambers2024}{}{}. In these objects, GI does not play a dominant role, and material is piled up in the inner regions of the dead zone by hydrodynamic instabilities \citep[e.g., ][]{Lyra2019, Cui2022}{}{} until the critical surface density for MRI activation is reached and the TI is initiated. \\
Simple 1D models by \cite{Chambers2024} indicate that the heating fronts launched by the TI create pressure maxima in the radial disk structure within 1 AU, which could serve as traps for drifting pebbles. The silicate sublimation front \citep[e.g., ][]{Kama2009}{}{} plays an important role in the initiation and development of the TI because the associated strong change in opacity constitutes a respective change in disk temperature by absorbing the bulk of the stellar irradiation \citep[][]{Flock2016, Flock2019}{}{}. Therefore, at least in passively heated disks, the location of the MRI transition and the location of the inner dust rim are closely related. \cite{Schobert2019} showed that the additional consideration of heating by viscous dissipation can shift the inner edge of the dead zone further outward while leaving the location of the dust sublimation front mostly unchanged for large accretion rates. However, analogously to \cite{Flock2019}, the models of \cite{Schobert2019} simulate the hydrostatic structure of the inner disk and do not consider the implications for the dynamical evolution. \\
Instead of focusing a priori on the recreation of observed luminosity features of outbursting T Tauri stars, this work aims to analyse the inner disk's stability and the consequences of potential TI-induced accretion events on the vertical and radial structure and evolution of a disk around a Class II T Tauri star. In our models, the TI is induced by the activation of the MRI in the dead zone. The `classical' TI caused by hydrogen ionisation and the consequent strong temperature dependence of the opacity does not occur in our simulations. In contrast to previous works investigating the hydrodynamic evolution, our models include a permanently MRI-active inner disk and the transition to the dead zone within the computational domain. Furthermore, we include heating by both stellar irradiation and viscous dissipation, radiation transport in the radial and vertical direction, and tracking of the variable dust-to-gas mass ratio throughout the inner 10 AU of the disk. This allows us to also monitor the evolution of the dust sublimation region during the TI phase. We take the hydrostatic model of the inner disk rim around a T Tauri star by \cite{Flock2019} as an initial model and evolve it in time under the influence of viscous heating. \\
The paper is organised as follows: In Sec. \ref{sec:method}, we describe the numerical and physical setup of our model, including the equations and parameters governing both the initial configuration and the time-dependent evolution. In Sec. \ref{sec:results}, the results of our simulations are presented and consecutively analysed and discussed in Sec. \ref{sec:discussion}. Finally, we summarize the main conclusions in Sec. \ref{sec:conclusion}.

\section{Method}\label{sec:method}
The simulations conducted in this work consist of two parts. First, an initial model was created by finding a solution for the hydrostatic equilibrium, including irradiation by the central star, dust sublimation and a simplified description of viscous heating. The creation of this model is elaborated on in Sec. \ref{sec:init}. Second, the hydrostatic solution was used as a starting configuration for the time-dependent radiation hydrodynamic simulation, which is described in Sec. \ref{sec:hydrodyn}. For both steps, we used the finite volume method of the PLUTO Code \citep[][]{Mignone2007}{}{}. In Sec. \ref{sec:viscosity}, we describe the implementation of turbulent viscosity in our model, while Sec. \ref{sec:dust_sublim} and Sec. \ref{sec:opacities} contain the description of dust sublimation and utilized opacities, respectively. Finally, in Sec. \ref{sec:numerica}, we give a brief overview of the numerical setup of the simulated domain together with the implemented boundary conditions.

\subsection{Initial model} \label{sec:init}

The construction of the hydrostatic initial model generally followed the description in \cite{Flock2019} with the addition of the effect of viscous heat dissipation. After building an initial surface density profile, depending on the viscosity $\nu$ and a chosen radially constant mass flux $\dot{M}_\mathrm{init}$ according to,

\begin{equation} \label{eq:Sigma}
    \Sigma=\frac{\dot{M}_\mathrm{init}}{3\, \pi \, \nu} ~,
\end{equation} 

\noindent and determining an initial temperature field $T(r, \theta)$ utilizing the optically thin solution, we solved the equations of hydrostatic equilibrium in spherical coordinates $(r, \theta, \phi)$ assuming axisymmetry. The solution provides a density- and azimuthal velocity field,  $\rho(r, \theta)$ and $v_\phi (r, \theta)$. Using the equation of state of an ideal gas,

\begin{equation}
    P=\frac{\rho \, k_\mathrm{B} \, T}{\mu_\mathrm{g} \, u} \; , \label{eq:state}
\end{equation}

\noindent to couple the thermal pressure $P$ with the temperature $T$, where $k_\mathrm{B}$ is the Boltzmann constant, $\mu_\mathrm{g}$ is the mean molecular weight and $u$ is the atomic mass unit, in conjunction with $\rho(r, \theta)$ we could find a steady-state solution of the system of two coupled radiative transfer equations, 

\begin{align}
    \frac{1}{\gamma -1} \, \pder[P]{t}=-\kappa_\mathrm{P} \, \rho_\mathrm{dust} \, c \, (a_\mathrm{R} \, T^4 - E_\mathrm{R}) - \nabla \cdot F_* + Q_\mathrm{acc} \; , \label{eq:rad1} \\
    \pder[E_\mathrm{R}]{t} - \nabla \frac{c \, \lambda}{\kappa_\mathrm{R} \, \rho_\mathrm{dust}} \, \nabla E_\mathrm{R} = \kappa_\mathrm{P} \, \rho_\mathrm{dust} \, c \, (a_\mathrm{R} \, T^4 - E_\mathrm{R}) \;,  \label{eq:rad2}
\end{align}

\noindent to obtain the internal energy and radiation field in radiative equilibrium. The equations above describe the heating, cooling and radiative diffusion in the flux-limited approximation, where $\gamma$ is the adiabatic index, $\kappa_\mathrm{P}$ and $\kappa_\mathrm{R}$ are the Planck- and Rosseland mean opacity, respectively, $a_\mathrm{R}=4 \, \sigma_\mathrm{B}/c$ is the radiation constant (with $\sigma_\mathrm{B}$ being the Stefan-Boltzmann-constant and $c$ being the speed of light), $E_\mathrm{R}$ is the radiation energy and $\lambda$ is the flux limiter function \citep{Levermore1981}. We assumed a blackbody irradiation flux $F_*$, which can be evaluated at every radius by,

\begin{equation}
    F_*(r)=\left ( \frac{R_*}{r} \right) ^2\, \sigma_\mathrm{B}\; T_*^4\; e^{-\tau_*} \; ,
\end{equation}

\noindent with consideration of the effect of the radial optical depth $\tau_*$ (which is described in Sec. \ref{sec:dust_sublim}). $R_*$ and $T_*$ are the stellar radius and the stellar surface temperature, for which we adopted the values of $R_*=2.6\, R_\odot$ and $T_*=4300\, K$, respectively, following the model setup of \cite{Flock2019}.\\
$Q_\mathrm{acc}$ describes the heating by viscous dissipation. Its implementation followed the procedure described in \cite{Schobert2019}. In hydrostatic equilibrium, it is assumed that the dominant contribution to the velocity field is $v_\phi$. Therefore, the viscous heating term in spherical coordinates takes the simplified form,

\begin{equation}
    Q_\mathrm{acc}=\mu \, \left [r \, \frac{\partial}{\partial r} \left (\frac{v_\phi}{r} \right ) \right ]^2 =\mu \, \left [r \, \pder[\Omega]{r} \right ]^2 \; , \label{eq_Qacc}
\end{equation}

\noindent with the orbital frequency $\Omega$ and dynamic viscosity $\mu$ as the product of the local  density $\rho$ and the kinematic viscosity $\nu$,

\begin{equation}
    \mu=\rho \, \nu = \rho \frac{\alpha c_\mathrm{s}^2}{\Omega} \; . \label{eq:viscosity}
\end{equation}

For the kinematic viscosity, we used the description introduced by \cite{Shakura1973} with the stress-to-pressure ratio $\alpha$ and the local speed of sound $c_\mathrm{s}$. \\
We note that if viscous heating and dust evaporation are considered and the mass accretion rate is high enough to enable significant heating by viscous dissipation, thermal instability is expected in the inner disk \citep[][]{Pavlyuchenkov2023}{}{}. This impedes the convergence of our computational method to find a hydrostatic solution for the initial model, which has also been recognised by \cite{Schobert2019}. This is why $\alpha$ was kept at a fixed value of $10^{-3}$ in the viscous heating term in Eq. \ref{eq:rad1} \citep[similar to ][]{Schobert2019}{}{} during the creation of the initial model. For determining the initial surface density distribution (Eq. \ref{eq:Sigma}) and during the hydrodynamic simulation, $\alpha$ was adapted according to the description given in Sec. \ref{sec:viscosity}. \\
The initial mass in the computational domain of our simulations was controlled by the radially constant mass flux $\dot{M}_\mathrm{init}$. Consequently, this also influences the pile-up of mass at the inner dead zone edge and, therefore, the effectiveness of viscous heating and heat-trapping, which potentially leads to the onset of the TI-induced accretion event. Furthermore, if an accretion event occurs, $\dot{M}_\mathrm{init}$ determines the available mass to be accreted and consequently has a significant influence on the duration and peak accretion rate of the burst. In this work, we implemented four values for $\dot{M}_\mathrm{init}$ ranging from $3.6\cdot 10^{-10}\, \mathrm{M}_\odot \, \mathrm{yr}^{-1}$ to $1\cdot 10^{-8}\, \mathrm{M}_\odot \, \mathrm{yr}^{-1}$ which is consistent with observed values in the young ($\leq$ 1 Myr) star cluster NGC 1333 \citep[][]{Fiorellino2021}{}{} as well as in the older star forming regions Lupus \citep[][]{Alcala2017}{}{} and Chameleon I \citep[][]{Manara2017}{}{}.

\subsection{Time-dependent simulation} \label{sec:hydrodyn}

For the time-dependent simulation, we solved the following set of radiation-hydrodynamic equations, 

\begin{align}
    &\pder[\rho]{t} \, + \nabla \cdot ( \rho \, \vec v ) = 0\;, \label{eq:cont}&& \\
    &\pder[\rho \, \vec v]{t} + \nabla \cdot (\rho \, \vec v  \, \vec v^T) + \nabla P_\mathrm{gas} = -\rho \, \nabla \Phi + \nabla \vec \Pi \;, \label{eq:mot} &&
\end{align}
\begin{multline}
    \pder[E]{t} + \nabla \cdot [(E + P_\mathrm{gas}) \, \vec v] = -\rho \, \vec v \cdot \nabla \Phi - \vec \Pi : \nabla \vec v  \\ 
    - \kappa_\mathrm{P} \, \rho \, c \, (a_\mathrm{R}  \, T^4 - E_\mathrm{R}) - \nabla \cdot F_*  \;, \label{eq:ene} 
\end{multline}

\noindent with the velocity vector $\vec v=(v_\mathrm{r}, v_\theta, v_\phi)$, the gas pressure $P_\mathrm{gas}$, the total energy $E$ and the gravitational potential $\Phi=G \, M_* /r$, where $G$ is the gravitational constant and $M_*$ is the mass of the star, which we fixed at $M_*=1\, M_\odot$ for this work. $\vec \Pi$ represents the viscous stress tensor, describing the viscous angular momentum transport in the equations of motion (Eq. \ref{eq:mot}) and viscous heating in the energy equation (Eq. \ref{eq:ene}) in the respective terms.
$\nabla \vec v$ (an abbreviation of $\nabla \otimes \vec v$ where `$\otimes$' is the dyadic product) is the velocity gradient tensor. Utilizing the identity $\vec A : (\vec b \otimes \vec c) = \vec b \cdot (\vec A \cdot \vec c)$, where $\vec A$ is a tensor and $\vec b$ and $\vec c$ are vectors, the viscous heating term in Eq. \ref{eq:ene} can also be written as,

\begin{equation}
    \vec \Pi : \nabla \vec v= \nabla \cdot (\vec \Pi \cdot \vec v) \; \label{eq:vischeating}
\end{equation}

\noindent with the stress tensor taking the form,

\begin{equation}
    \vec \Pi=\mu \left [ \nabla \vec v + (\nabla \vec v)^\mathrm{T}-\frac{2}{3}(\nabla \cdot \vec v)\vec {\mathrm{I}} \right ] \; .
\end{equation}

\noindent $\vec {\mathrm{I}}$ denotes the unit tensor. If the contributions of the radial and polar components of the velocity can be neglected, Eq. \ref{eq:vischeating} results in the expression for $Q_\mathrm{acc}$ (Eq. \ref{eq_Qacc}). \\
As a closure relation, we again used the ideal gas equation. After finding a solution to the system of Eqs. \ref{eq:cont}, \ref{eq:mot} and \ref{eq:ene}, the coupled equations for the radiation transport, Eqs. \ref{eq:rad1} and \ref{eq:rad2}, were solved to acquire the temperature and the radiation energy for the next time step. Since the heating by viscous dissipation was already accounted for by the corresponding term in the total energy equation (where the contributions from the radial and polar velocity gradients are considered as well), we omitted the viscous heating term $Q_\mathrm{acc}$ in Eq. \ref{eq:rad1} during the time-dependent simulation.\\

\subsection{Viscosity} \label{sec:viscosity}

The stress-to-pressure ratio entering the viscous stress description can adopt two different values, $\alpha_\mathrm{MRI}$ or $\alpha_\mathrm{DZ}$, depending on the local temperature. $\alpha_\mathrm{MRI}$ was implemented if the temperature of the environment was high enough to sustain MRI, while $\alpha_\mathrm{DZ}$ was used at lower temperatures (in the dead zone). To ensure a smooth transition between these values around a critical temperature $T_\mathrm{MRI}$, the stress-to-pressure ratio takes the form,

\begin{equation}
    \alpha=(\alpha_\mathrm{MRI}-\alpha_\mathrm{DZ}) \frac{1}{2} \left [1 - \mathrm{tanh}\left ( \frac{T_\mathrm{MRI}-T}{25\mathrm{K}}\right)\right]+\alpha_\mathrm{DZ} \; . \label{eq_alpha}
\end{equation}

In our simulations, $\alpha_\mathrm{MRI}$ was set to a value of 0.1. This is motivated by the findings of \cite{Flock2017}, who argue that the stress-to-pressure ratio can be as high as 10\% in the MRI active zone if a net vertical flux is present. Furthermore, models with $\alpha_\mathrm{MRI}=0.1$ have been shown to be consistent with observed signatures of YSOs undergoing accretion burst events \citep[e.g., ][]{Cleaver2023, Liu2022}. The residual viscosity in the dead zone $\alpha_\mathrm{DZ}$ was chosen to be $10^{-3}$, which represents hydrodynamic turbulence \citep[e.g., ][]{Lyra2019, Cui2022}{}{}. This value also ensures that the surface density in the computational domain is small enough to keep the Toomre parameter \citep{Toomre1964} well above 1. Therefore, we did not consider the effects of gravitational instability in our simulations. In accordance with \cite{Flock2019}, we assumed the MRI activation temperature $T_\mathrm{MRI}$ to be 900\,K. \\

\subsection{Dust sublimation} \label{sec:dust_sublim}

Following \cite{Flock2016} and \cite{Flock2019}, we used the following fitted function for the evaluation of the dust sublimation temperature \citep[][]{Isella2005}{}{},
\begin{equation}
    T_\mathrm{S}=2000\mathrm{K} \left( \frac{\rho}{1 \, \mathrm{g \, cm^{-3}}}\right)^{0.0195} \; , \label{eq:T_S}
\end{equation}
which then allowed us to determine the dust-to-gas mass ratio,

\begin{equation}
    f_\mathrm{D2G}=
    \begin{cases}
    f_{\Delta \tau} \frac{1}{4} \left [ 1- \mathrm{tanh}\left ( \left (\frac{T-T_\mathrm{S}}{150 \, \mathrm{K}} \right )^3 \right)\right] \left [ 1- \mathrm{tanh}(2/3-\tau_*)\right ] & \\
    & \\
    f_0 ~ \text{for $T<T_\mathrm{S}$ and $\tau_*>3.0$}
    \end{cases} .\label{eq:fd2g}
\end{equation}

$f_{\Delta \tau}$ describes the respective dust-to-gas mass ratio that results in an optical depth of $\Delta \tau=0.2$ in a computational grid cell. It takes the form $f_{\Delta \tau}=0.2/[\rho \, \kappa_\mathrm{P}(T_*) \, \Delta r]-\kappa_\mathrm{gas}/\kappa_\mathrm{P}(T_*)$, where $\kappa_\mathrm{P}(T_*)$ is the dust opacity resulting from averaging the wavelength-dependent opacity over the spectrum of the light from the star irradiating the disk, $\kappa_\mathrm{gas}$ is the gas opacity and $\Delta r$ is the radial extent of the grid cell. $f_{\Delta \tau}$ ensures that the absorption of radiation by a single cell is limited, allowing us to properly resolve the absorption at the inner edge of the dust disk. Eq. \ref{eq:fd2g} describes a smooth transition of the dust-to-gas ratio between a minimum value of $10^{-10}$ (which was set to facilitate numerical stability) and the maximum value of $f_0$ around the sublimation temperature\footnote{The power of 3 in the hyperbolical tangent function is not strictly necessary, but was kept to stay consistent with the description of \cite{Flock2019}} and around the area where a radial optical depth of $\tau_*=2/3$ is reached, which is where most of the stellar irradiation is absorbed. In this context, the radial optical depth was calculated as,

\begin{equation}
    \tau_*=\tau_0 + \int_{r_0}^{r} \sigma_* ~dr \; , \label{eq:tau}
\end{equation}

\noindent with $\tau_0=\rho_{r_0} \, \kappa_\mathrm{gas} \, (r_0-R_*)$ being the optical depth at the inner radius of the computational domain $r_0$, assuming that the dust-to-gas ratio in the disk between $r_0$ and the stellar surface is zero. $\sigma_*$ was given by the sum of the contributions to the optical depth from dust and gas, $\sigma_*=\rho_\mathrm{dust} \, \kappa_\mathrm{P}(T_*) + \rho_\mathrm{gas} \, \kappa_\mathrm{gas}$. If $\tau_*>3.0$ and the temperature of the disk material is smaller than the local sublimation temperature, we set the dust-to-gas ratio to its maximum value $f_0$. We adopted $f_0=10^{-3}$ to account for the part of the dust content that contributes to the opacity used in the radiative transfer scheme and is well-coupled to the gas. $f_0$ does not include grown dust already settled towards the midplane, which does not play a significant role in determining the infrared opacity.

\subsection{Opacities} \label{sec:opacities}
As described in \cite{Flock2019}, $\kappa_\mathrm{gas}$ was taken as an average of Rosseland mean opacities calculated by \cite{Malygin2014}. The average was calculated for the parameter space relevant to our models. \cite{Flock2019} derived an average value of $\kappa_\mathrm{gas}=10^{-5}\, \mathrm{cm}^{2}\, \mathrm{g}^{-1}$ for the conditions of the gas phase in the hydrostatic case. The same value has been adopted here for the creation of the initial model. During the hydrodynamic simulation, we recalculated the mean value for the gas opacity to account for the increased temperature, density and pressure of the gas disk during phases of thermal instability. By considering this adapted parameter space, we derived a value of $\kappa_\mathrm{gas}=10^{-3}\, \mathrm{cm}^{2}\, \mathrm{g}^{-1}$ for the Rosseland mean gas opacity (see Appendix \ref{sec:app_opac}). Following \cite{Flock2019}, we assumed the same value for the irradiation opacity. 
With this increase in gas opacity, optically extremely thin regions could be avoided. Such regions are unable to cool efficiently, which leads to strong temperature increases when heating mechanisms (in addition to irradiation heating) in the time-dependent simulations are considered. As a consequence, the numerical stability of the hydrodynamic simulations is significantly enhanced. 
For the dust opacities, we assumed $\kappa_\mathrm{P}=\kappa_\mathrm{R}$ and adopted the two values derived by \cite{Flock2019}: the irradiation opacity corresponding to the stellar surface temperature, $\kappa_\mathrm{P}(T_*)=1300\, \mathrm{cm}^{2}\, \mathrm{g}^{-1}$, and the thermal emission opacity corresponding to a typical dust sublimation front temperature $\kappa_\mathrm{P}(1500\, \mathrm{K})=700\, \mathrm{cm}^{2}\, \mathrm{g}^{-1}$.

\subsection{Numerical domain configuration and boundary conditions} \label{sec:numerica}
The setup of the computational domain was the same as in \cite{Flock2019}.
The simulations incorporated the inner part of a protoplanetary disk in spherical coordinates with the radial extent ranging from $r_\mathrm{in}=0.05$\,AU to $r_\mathrm{out}=10$\,AU and the meridional extent being symmetrical around the disk midplane with $\theta=\pi/2 \pm 0.15$. The numerical grid consisted of $N_\mathrm{r}=2048$ logarithmically spaced grid points in the radial direction and $N_\theta=128$ linearly spaced points in the meridional direction.\\
We imposed Neumann (zero gradient) conditions at the radial inner ($r_\mathrm{in}$) and outer ($r_\mathrm{out}$) boundary for the radial and poloidal velocity. A linear extrapolation was used for the azimuthal velocity and the density to facilitate the continuation of the gradients of the respective quantities. \\
At the upper and lower poloidal boundaries, the continuation of the density gradient was implemented as well while using free Neumann conditions for all velocity components. \\
An inflow of mass into the computational domain was undesired for our models. Therefore, if the zero gradient condition would allow for an inflow (i.e. if the velocity in the cells next to the boundary is pointing into the domain), the velocities were reflected in the boundary\footnote{E.g. if the radial velocity $v_\mathrm{r}$ is defined such that positive values correspond to a flow towards larger radii and $v_\mathrm{r}(r_\mathrm{out}, \theta)<0$, then $v_\mathrm{r}(\partial_r^+ \mathbf{U}, \theta)=-v_\mathrm{r}(r_\mathrm{out}, \theta)$ with $\mathbf{U}$ being the computational domain and $\partial_r^+ \mathbf{U} $ being its numerical outer radial boundary.}. Although we calculate the conditions in the disk beyond $\sim$ 2 AU in the same manner as in the inner disk, for the purpose of our simulations, these regions technically act as a mass reservoir for refilling the inner disk after it has been depleted by the TI-induced accretion event. Since the viscous timescale at the outer boundary (10 AU) is of the order of $10^6$ yr, which is much larger than the time frame simulated by our models, the chosen no-inflow condition at the outer radial boundary should not affect our results. Furthermore, global disk models have shown that TI-induced accretion events can still occur when the inner disk is cut off from resupply of material from radii greater than 10 AU by, e.g. gaps in the disk structure \citep[][]{Cecil2024}{}{}.\\
A zero gradient condition was used for the temperature at all four boundaries. The radiation energy density at the boundaries was determined by $E_\mathrm{R}(\partial \mathbf{U})=\sigma [T_\mathrm{0}(\partial \mathbf{U})]^4$ where $\partial \mathbf{U}$ is the boundary of the domain. Hereby, $T_\mathrm{0}$ is the temperature of a disk region that is optically thin with respect to the irradiation of starlight and is calculated as,

\begin{equation}
    T_\mathrm{0}=\frac{1}{2}\, \epsilon^{1/4}\, \left (\frac{R_*}{2\, r}\right )^{1/2}\, T_* \; ,
\end{equation}

\noindent which was also used as an initial temperature field for creating the hydrostatic initial model. $\epsilon$ is the ratio between the absorption and emission efficiencies, which simplifies in this case to the ratio of irradiation opacity $\kappa_\mathrm{P}(T_*)$ to thermal emission opacity $\kappa_\mathrm{P}(T_\mathrm{S})$ when neglecting the contribution of the gas opacity \citep[e.g.,][]{Ueda2017}{}{}. The factor $1/2$ was included to allow for efficient cooling of the disk by radiation\footnote{We also tested a lower value of this factor of 0.05 and observed no significant impact on the hydrodynamic evolution.}.

\section{Results}\label{sec:results}

The purpose of this work is to analyse the temporal behaviour of the inner parts of a protoplanetary disk under the influence of stellar irradiation, radial and vertical radiative transfer, dust sublimation, turbulent $\alpha$-parameterized activity and viscous heating. For that purpose, we constructed a reference model \texttt{MREF}, the parameters of which are summarized in Tab. \ref{tab:MREF}. Several additional models were set up to investigate the influence of different initial accretion rates and, consequently, different inner disk masses. An overview of the chosen parameters and names for the different models is given in Tab. \ref{tab:models}.

\begin{table}[t]
{\renewcommand{\arraystretch}{1.5}
\caption{Implemented parameters for the reference model \texttt{MREF}.}
\begin{tabular}{lr|lr}
\hhline{====}
$M_*$ {[}$\mathrm{M}_\odot${]}                           & 1.0                & $\kappa_\mathrm{P}(T_*)$ {[}$\mathrm{cm}^2\, \mathrm{g}^{-1}${]}          & 1300              \\
$R_*$ {[}$\mathrm{R}_\odot${]}                           & 2.6                & $\kappa_\mathrm{P}(T_\mathrm{S})$ {[}$\mathrm{cm}^2\, \mathrm{g}^{-1}${]} & 700               \\
$T_*$ {[}K{]}                                            & 4300               & $\kappa_\mathrm{gas}$ {[}$\mathrm{cm}^2\, \mathrm{g}^{-1}${]}             & $10^{-3}$         \\
$\alpha_\mathrm{MRI}$                                    & 0.1                & $r_\mathrm{in}$ {[}AU{]}                                                  & 0.05              \\
$\alpha_\mathrm{DZ}$                                     & $10^{-3}$          & $r_\mathrm{out}$ {[}AU{]}                                                 & 10                \\
$\dot{M}_\mathrm{init}$ {[}$\mathrm{M}_\odot \, \mathrm{yr}^{-1}${]} & $3.6\cdot 10^{-9}$ & $\theta$ {[}rad{]}                                                        & $\pi /2 \pm 0.15$ \\
$T_\mathrm{MRI}$ {[}K{]}                                 & 900                & $N_\mathrm{r}$                                                            & 2048              \\
$f_0$                                                    & $10^{-3}$          & $N_\theta$                                                                & 128 \\
\hline
\end{tabular}
\label{tab:MREF}
}
\end{table}

\begin{table}[t]
\caption{Initial accretion rates for the different models analysed in this work.}
\centering
\begin{tabular}{cc}
\hhline{==}
Model       & \begin{tabular}[c]{@{}c@{}}$\dot{M}_\mathrm{init}$\\ {[}$\mathrm{M}_\odot \, \mathrm{yr}^{-1}${]}\end{tabular}   \\ \hline
$\tt{MREF}$ & $3.6 \times 10^{-9}$                                                                                                                                                                                                                                     \\
$\tt{M1}$   & $3.6 \times 10^{-10}$                                                                                                                                                                                                                                  \\
$\tt{M2}$   & $1 \times 10^{-9}$                                                                                                                                                                                                                                       \\
$\tt{M3}$   & $1 \times 10^{-8}$                                                                                                                                              \\ \hline
\end{tabular}
\tablefoot{All other parameters listed in Tab. \ref{tab:MREF} are the same for all setups.}
\label{tab:models}
\end{table}

\subsection{Initial configuration} \label{sec:init_config}

\begin{figure*}[ht!]
    \centering
         \resizebox{\hsize}{!}{\includegraphics{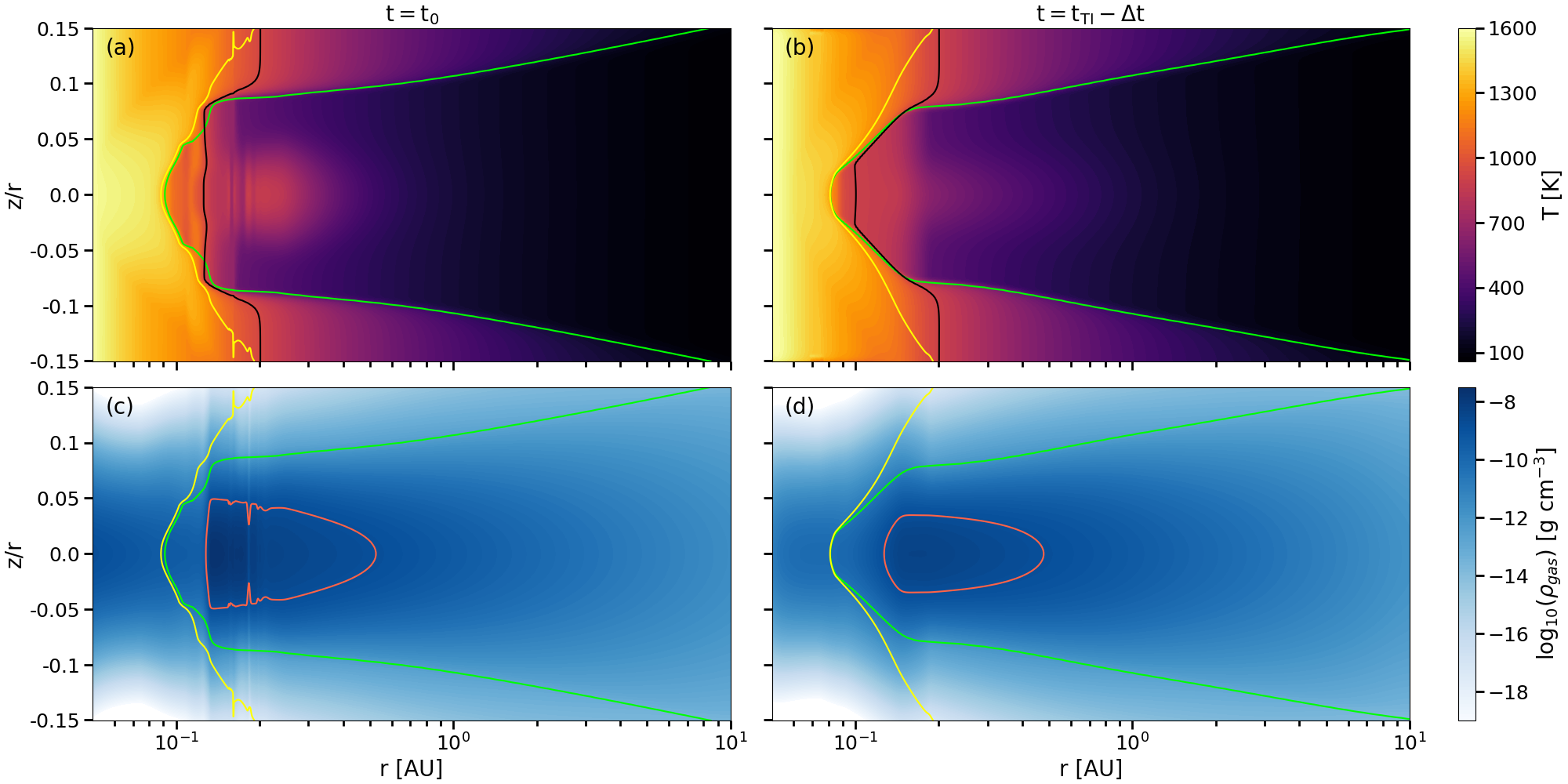}}
    \caption{
    Temperature (panels a and b) and gas density (panels c and d) maps for the model \texttt{MREF} in the initial configuration at $t=t_0$ and at the end of the quiescent phase, shortly before the thermal instability is triggered, $t=t_{TI}-\Delta t$. Yellow contour lines depict the region of dust sublimation ($T=T_\mathrm{S}$), black lines show the ionization transition from the MRI active zone to the dead zone ($T=T_\mathrm{MRI}$) and green lines mark the radial optical depth $\tau_*=2/3$ surface. In panels c and d, the red contours are drawn at $\rho_\mathrm{gas}=10^{-9}$ which corresponds to a dust density of $\rho_\mathrm{dust}=10^{-12}$.
    }
    \label{fig:init_quiescent}
\end{figure*}

\begin{figure}[ht!]
    \centering
         \resizebox{\hsize}{!}{\includegraphics{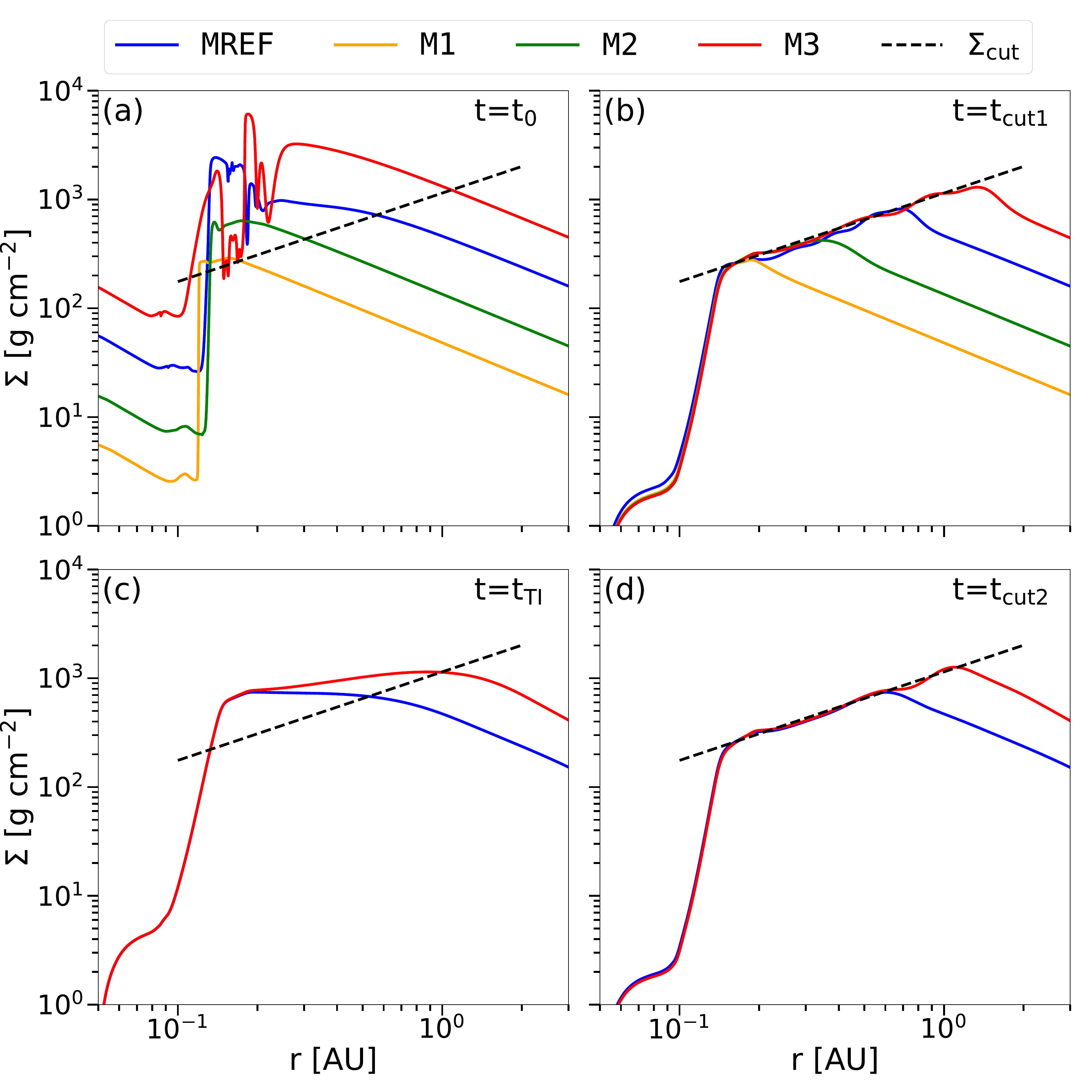}}
    \caption{Surface density profiles of models with different $\dot{M}_\mathrm{init}$ at distinct evolutionary stages. Panel a shows the initial configuration for all models, while panel b presents their structure after the first TI and the subsequent diffusion of the resulting density bumps. Panel c compares the surface density of \texttt{MREF} and \texttt{M3} at the start of their respective second TI cycle, and panel d shows the structure of both models after they have become quiescent again (similar to panel b). In all cases, the TI phase persists until enough mass has been accreted so that the surface density does not exceed $\Sigma_\mathrm{cut}$.}
    \label{fig:sigmacompare_mdots}
\end{figure}

Panels a and c of Fig. \ref{fig:init_quiescent} show the temperature and gas density map of the initial configuration of the reference model MREF. The structure of the disk is similar to the one described in \cite{Flock2016}, \cite{Flock2019} and \cite{Schobert2019}: A pure gas disk is present at radii smaller than 0.09 AU, followed by an arched inner dust rim between 0.09 AU and 0.15 AU. The transition from an MRI active inner disk to the dead zone occurs at a radius of 0.15 AU in the midplane. The transition region stays at a constant radius in the vertical direction until it crosses the $\tau_*=2/3$ line, after which the medium in the line of sight towards the star is optically thin, and the MRI transition moves outwards. After a shadowed region between $\sim$ 0.15 AU and $\sim$ 0.25 AU, a flared disk extends towards $r_\mathrm{out}$. At radii between 0.15 AU and 0.2 AU, which is the region where the MRI transition takes place, multiple vertical ripples are visible in the temperature and density distribution. These disturbances are an indication of thermal instability. The conditions in the initial configuration of the \texttt{MREF} model are such that viscous heating and heat-trapping around the disk midplane shortly beyond the MRI transition are effective enough to elevate the temperature to values at which the viscous $\alpha$ parameter becomes dependent on the temperature (according to Eq. \ref{eq_alpha}). However, the development of the TI is suppressed by the computational method, which aims to find a hydrostatic solution. In fact, we found multiple solutions for the hydrostatic structure, between which the solving algorithm switches after every iteration. This finding is similar to the conclusion of \cite{Pavlyuchenkov2023}, who find that there is no unique solution to the thermal structure when considering accretion heating, dust evaporation, effects of gas opacity and a sufficiently high accretion rate. In the context of S-curves of thermal balance, this means that the surface density adopts values which allow for at least two temperature values, all of which would result in a vertical thermal equilibrium. However, we found that the choice of the initial model between these different configurations is irrelevant to the evolution of the disk during the hydrodynamic simulation. Although the stress-to-pressure ratio $\alpha$ is kept constant in the calculation of the viscous heating term for the hydrostatic equilibrium, the surface density is determined according to Eq. \ref{eq:Sigma} with the contribution of $\alpha$ as described in Eq. \ref{eq_alpha}. A change in temperature in the vicinity of $T_\mathrm{MRI}$ due to viscous heating translates to a change in $\alpha$ and consequently to an alteration of $\Sigma$. Therefore, the non-existence of a unique solution for the thermal structure manifests itself in the form of jumps in $\Sigma$, leading to the ripples visible in panels a and c of Fig. \ref{fig:init_quiescent}. During the hydrodynamic simulation, these ripples are smoothed out and the disk adopts the structure visible in panels b and d of Fig. \ref{fig:init_quiescent} unless 
 a thermal instability is in progress (which is analysed in Sec. \ref{sec:TI_phase}).
 Fig. \ref{fig:sigmacompare_mdots} shows the surface density distributions of the models listed in Tab. \ref{tab:models} at various stages during their evolution. The profiles depicted in panel a depict the initial configuration (at $t=t_0$).  The unstable region of \texttt{MREF} (blue line) is discernible here as well by the jumps in $\Sigma$ due to the aforementioned reasons. In the model \texttt{M3} with a higher initial accretion rate, this unstable behaviour is magnified due to the increased viscous heating caused by a larger density in the dead zone. While the ripples in the surface density of \texttt{M2} are less severe, they are almost completely absent in \texttt{M1} because the viscous heat dissipation and heat-trapping close to the midplane are less effective in these less massive disks.

\subsection{Thermal instability phase}\label{sec:TI_phase}

\begin{figure*}[ht!]
    \centering
         \resizebox{\hsize}{!}{\includegraphics{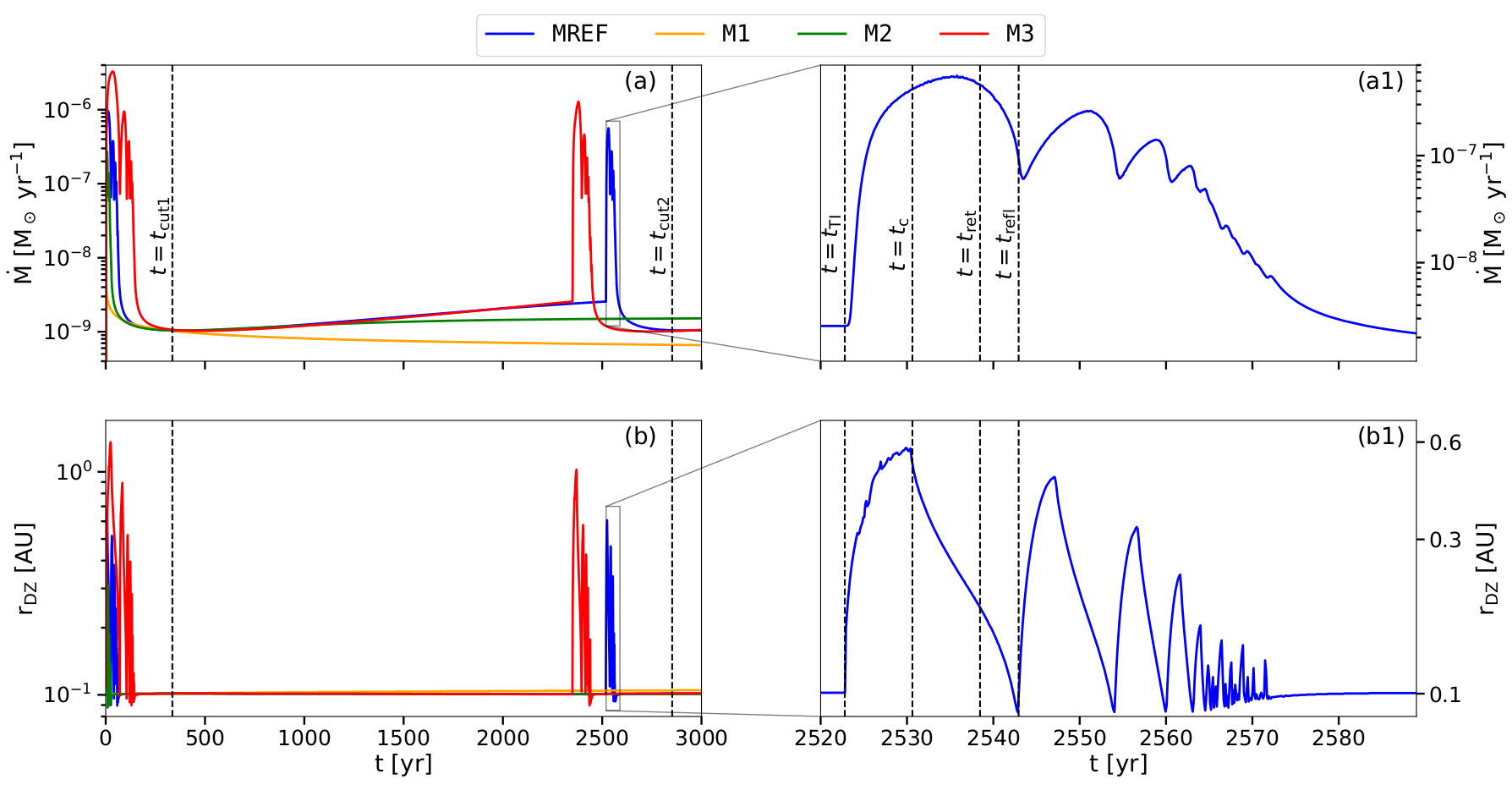}}
    \caption{Evolution of the accretion rate (panel a) and the position of the dead zone inner edge at the midplane (panel b) for the four different models analysed in this work. Panels a1 and b1 show a zoom-in to the time frame in which the TI-induced accretion event is occurring in the \texttt{MREF} model. The vertical dashed lines in panel a1 indicate the times corresponding to the ignition of the TI ($t_\mathrm{TI}$), the heating front reaching its largest extent and a cooling front starting to develop ($t_\mathrm{c}$), the retreat of the cooling front towards the star ($t_\mathrm{ret}$) and the ignition of the first reflare ($t_\mathrm{refl}$). In panel a, the vertical dashed lines mark the points in time for which the surface profiles of the \texttt{MREF} model in panels b ($t_\mathrm{cut1}$) and d ($t_\mathrm{cut2}$) of Fig. \ref{fig:sigmacompare_mdots} are shown.} 
    \label{fig:accr_rate_inset}
\end{figure*}

After starting the hydrodynamic simulation, \texttt{MREF} as well as \texttt{M2} and \texttt{M3} immediately enter the TI stage, which is expected since the initial models already show signs of instability. We acknowledge that this first dynamic instability phase may be strongly influenced by our choice of the hydrostatic initial configuration and the transition of the gas opacity from $10^{-5}\, \mathrm{cm}^2\, \mathrm{g}^{-1}$ in the hydrostatic case to $10^{-3}\, \mathrm{cm}^2\, \mathrm{g}^{-1}$ in the hydrodynamic case. Therefore, we refrain from conducting a detailed analysis of the initial TI cycles. Model \texttt{M1} does not show the development of TI since the density is small enough so that the combination of a low level of viscous heat dissipation and the small vertical optical depth does not lead to efficient heat-trapping inside the disk. 

The models' surface densities after the initial burst (at a time denoted as $t_\mathrm{cut1}$) are shown in panel b of Fig. \ref{fig:sigmacompare_mdots}. We find that in all cases, the amount of material accreted onto the star is such that the surface density after the TI phase smoothes out into a structure that does not exceed a characteristic profile $\Sigma_\mathrm{cut}\propto r^{0.8137}$. In panel a of Fig. \ref{fig:sigmacompare_mdots}, $\Sigma_\mathrm{cut}$ is plotted over the surface densities in the initial configuration. Notably, all models with surface densities larger than $\Sigma_\mathrm{cut}$ enter the outbursting stage. The small amount of mass which causes the surface density of \texttt{M1} to barely rise above $\Sigma_\mathrm{cut}$ around 0.15~AU is rapidly accreted onto the star without similar TI properties compared to the other models. \\
The models \texttt{MREF} and \texttt{M3} enter another TI-induced outbursting phase at a time referred to as $t_\mathrm{TI}$. The surface densities of those two models at their respective $t=t_\mathrm{TI}$ are shown in panel c of Fig. \ref{fig:sigmacompare_mdots}. $\Sigma_\mathrm{cut}$ is depicted again as a reference. The comparison to panel b makes clear that the density maximum present shortly before and shortly beyond 1 AU, respectively, after the initial instability (at $t=t_\mathrm{cut1}$), has smoothed out during the quiescent phase between the TI cycles and mass has been accreted towards the star, increasing the surface density above $\Sigma_\mathrm{cut}$ again. The mass tends to accumulate at the inner edge of the dead zone due to the large amount of angular momentum transported outwards from the highly viscous, MRI-active inner disk. As soon as enough material has accumulated so that viscous heating near the midplane is effective enough and the vertical optical depth from the midplane towards the disk surface is large enough to trap the heat and increase the midplane temperature towards values in the vicinity of $T_\mathrm{MRI}$, the TI develops and the disk enters the outbursting stage. The mass accumulation during the quiescent phase in the models \texttt{M1} and \texttt{M2} is insufficient for triggering a TI-cycle. Panel d of Fig. \ref{fig:sigmacompare_mdots} shows the surface density profiles of \texttt{MREF} and \texttt{M3} after their respective second TI phases (at $t=t_\mathrm{cut2}$). Analogously to the disk structures shown in panel b, enough mass has been accreted so that the surface density does not exceed $\Sigma_\mathrm{cut}$. 

Panels a and b of Fig. \ref{fig:accr_rate_inset} show the evolution of the accretion rate onto the star $\dot M$ and of the radius of the inner edge of the dead zone at the midplane $r_\mathrm{DZ}$ over a time-frame of 3000~yr for the models with varying initial accretion rates. After the initial bursts at the start of the simulations (in all models except for \texttt{M1}), the disks enter a quiescent stage in which the inner disk is resupplied with material accreted from larger radii.
The second TI cycles occurring in the models \texttt{MREF} and \texttt{M3} are initiated at times of 2522~yr and 2350~yr, respectively, after the start of the hydrodynamic simulation. The thermal- and density structure of \texttt{MREF} at the beginning of this phase (shortly before the TI is triggered, denoted with $t=t_\mathrm{TI}-\Delta t$) is depicted in panels b and d of Fig. \ref{fig:init_quiescent}. The shape of the inner rim is similar to the one of the initial model described in Sec. \ref{sec:init_config} with the notable differences that the disk is thinner and stretches further towards the star. Furthermore, the disk in panels b and d is the solution to the full set of radiation hydrodynamic equations, which means that potential ripples in the density structure are either smoothed out or lead to a runaway TI in contrast to the hydrostatic case shown in panels a and c. The state of the disk in the quiescent phase and just before the onset of the TI is very similar to the structure of the inner disk shown in Fig. 1. of \cite{Flock2019}. The viscous heating effect in our model manifests in the slightly larger contrast between the midplane temperature and the temperatures between the midplane and the $\tau_*=2/3$ surface. 

\begin{figure*}[ht!]
    \centering
         \resizebox{\hsize}{!}{\includegraphics{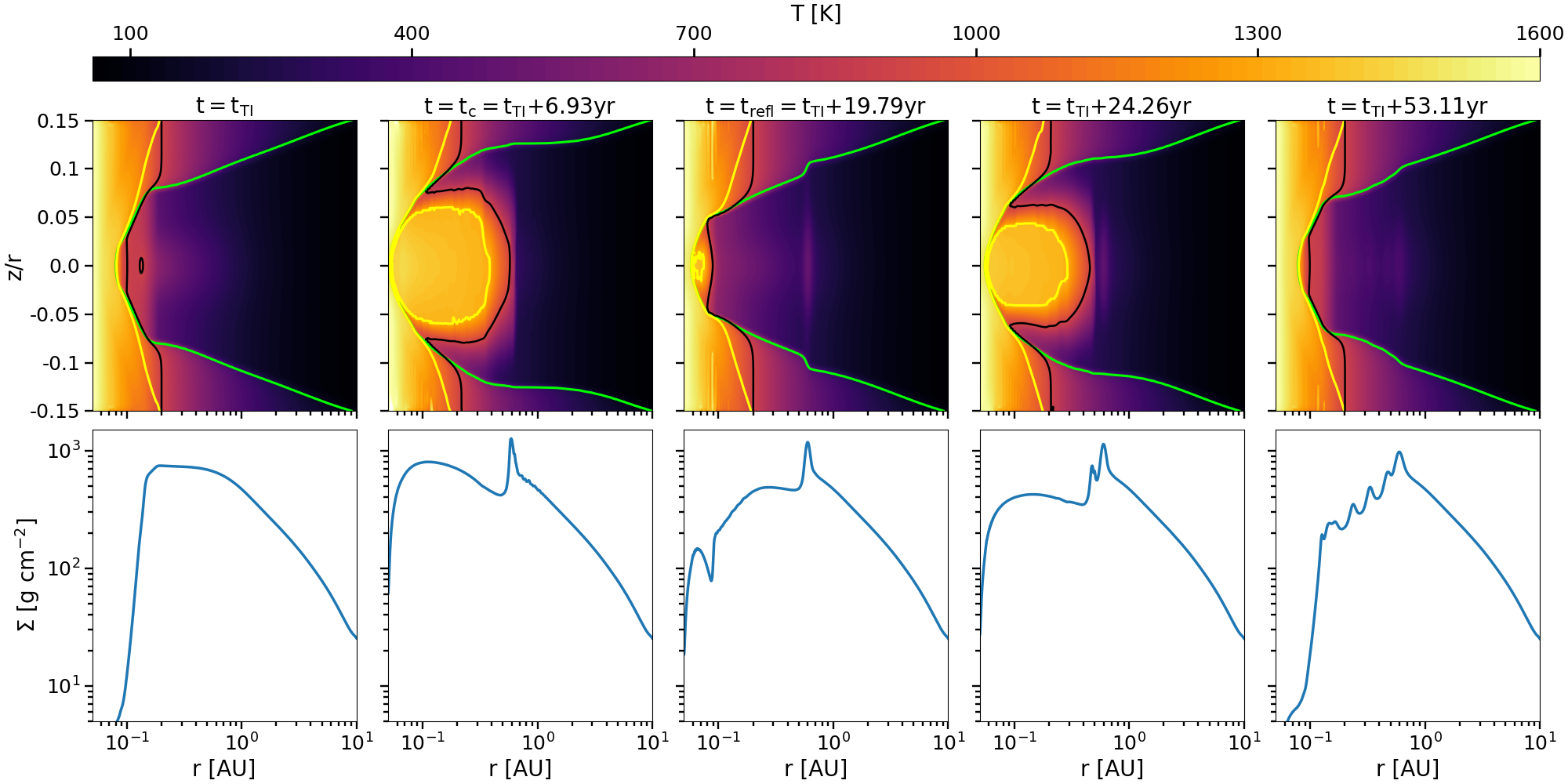}}
    \caption{
    Temperature maps (upper row) and corresponding surface density profiles (lower row) for four snapshots in time during an outburst event. In the upper row, the black contour lines correspond to $T=T_\mathrm{MRI}$, the yellow ones to $T=T_\mathrm{S}$ and the green ones to a radial optical depth of $\tau_*=2/3$. $t_\mathrm{TI}$ is chosen to be the point in time at which the thermal instability is triggered. The second column shows the state of the disk at the time the MRI-active region has reached its largest extent during the main burst ($t_\mathrm{c}$). The third column represents the stage of the ignition of the first reflare ($t_\mathrm{refl}$), the maximum extent of which is reached at the time depicted in column four. The state of the disk shortly after the instability has ended is shown in column five.
    }
    \label{fig:burststructure}
\end{figure*}

Panels a1 and b1 of Fig. \ref{fig:accr_rate_inset} are a zoom-in to the temporal evolution of the mass accretion rate and radius of the dead zone inner edge at the midplane, respectively, for the model \texttt{MREF} during the burst. The accretion event consists of multiple reflares, which decrease in amplitude with time. The vertical dashed lines in panels a1 and b1 indicate points in time, which are used to analyse the reflare mechanism in Sec. \ref{sec:S_curve_reflares}. Tab. \ref{tab:burst_props} summarizes several properties of the outbursts occurring in \texttt{MREF} and \texttt{M3}. We define the total duration of the burst $t_\mathrm{burst}$ as the time between the start of the rapid rise of the accretion rate and the point in time at which the accretion rate has decreased back to the pre-burst value. During this timeframe, a maximum accretion rate of $\dot{M}_\mathrm{max}$ is reached, which corresponds to an amplification of a factor $f_\mathrm{amp}$ with respect to the accretion rate just before the TI sets in. A total mass of $M_\mathrm{acc}$ is accreted onto the star over the course of an outburst and the inner edge of the dead zone at the midplane travels to a maximum distance of $r_\mathrm{DZ, max}$. The comparison with the values extracted from the accretion event in \texttt{M3} shows that all these quantities are universally increased in a more massive disk.

\begin{table}[]
\caption{Properties of the TI-induced accretion events occurring in the models \texttt{MREF} and \texttt{M3}. \texttt{M1} and \texttt{M2} do not accumulate enough mass to trigger the TI.}
\begin{tabular}{cccccc}
\hhline{======}
Model                          & \begin{tabular}[c]{@{}c@{}}$t_\mathrm{burst}$\\ {[}yr{]}\end{tabular} & \begin{tabular}[c]{@{}c@{}}$\dot{M}_\mathrm{max}$\\ {[}$\mathrm{M}_\odot~\mathrm{yr}^{-1}${]}\end{tabular} & $f_\mathrm{amp}$  & \begin{tabular}[c]{@{}c@{}}$M_\mathrm{acc}$\\ {[}$\mathrm{M}_\odot${]}\end{tabular} & \begin{tabular}[c]{@{}c@{}}$r_\mathrm{DZ, max}$\\ {[}AU{]}\end{tabular} \\ \hline
\texttt{MREF} & 54                                                          & $5.5\cdot 10^{-7}$                                                                                         & $2.2\cdot 10^{2}$ & $2\cdot 10^{-5}$                                                                    & 0.6                                                                     \\
\texttt{M3}   & 105                                                         & $1.3\cdot 10^{-6}$                                                                                         & $5.2\cdot 10^2$   & $8.5\cdot 10^{-5}$                                                                  & 1 \\  \hline                                                                      
\end{tabular}
\label{tab:burst_props}
\end{table}

\begin{figure}[h]
    \centering
         \resizebox{\hsize}{!}{\includegraphics{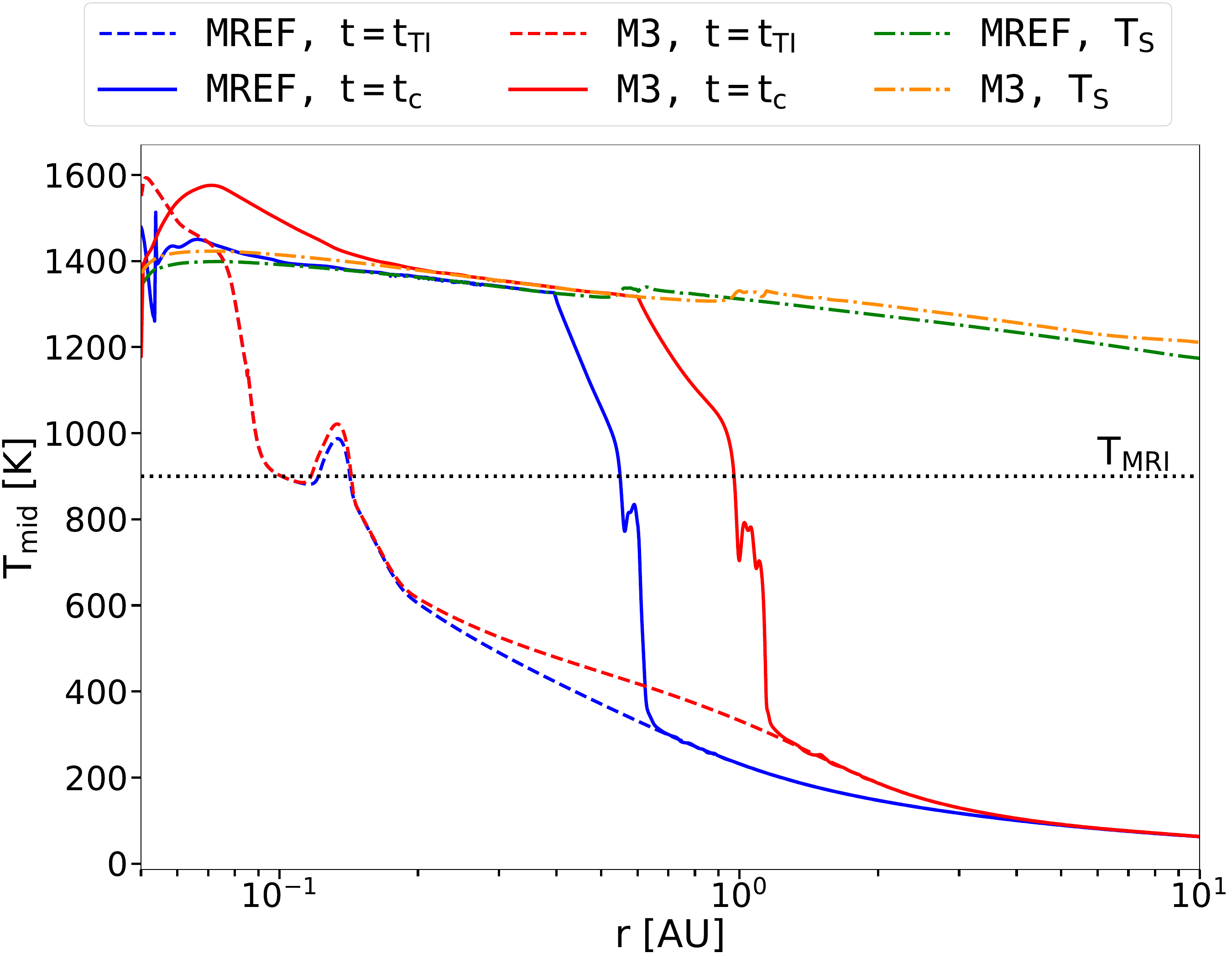}}
    \caption{
    Midplane temperature profiles of the \texttt{MREF} and \texttt{M3} models at the time of the onset of the TI ($t_\mathrm{TI}$) and at the time at which the cooling front develops ($t_\mathrm{c}$).
    }
    \label{fig:Tmid}
\end{figure}

Snapshots of the temperature structure and corresponding surface density at certain times during the burst in \texttt{MREF} are shown in Fig. \ref{fig:burststructure}. At $t=t_\mathrm{TI}$ (first column, corresponding to the stage of the disk shown in panel c of Fig. \ref{fig:sigmacompare_mdots}), the temperature in the dead zone is raised above $T_\mathrm{MRI}$ first at the midplane at a radius of 0.135~AU, which is slightly outside the radius of the midplane dead zone inner edge at 0.10~AU. The radial profile of the midplane temperature at $t=t_\mathrm{TI}$ for \texttt{MREF} is depicted as the blue dashed line in Fig. \ref{fig:Tmid} in comparison with the more massive \texttt{M3} model. In both models, the instability is ignited at the same position, which is a consequence of $r_\mathrm{DZ}$ being the same in both cases during the quiescent phase (panel b od Fig. \ref{fig:accr_rate_inset}).  The rise in $\alpha$ caused by the temperature increase leads to more rigorous viscous dissipation, letting the temperature build up rapidly further towards the dust evaporation temperature $T_\mathrm{S}$ at the point of first ignition. The material in the surrounding area is quickly heated up and becomes MRI-active as well. This launches an ionisation front in all directions, promptly reaching the inner permanently MRI-active gas disk. The ionisation (or heating) front is characterized as the moving boundary between the MRI-active and -inactive regions when it propagates into the dead zone. In \texttt{MREF}, the freshly ignited MRI-active zone reaches z/r=$\pm0.06$ within a time of about one month and continues to expand vertically as the ionisation front travels outwards. After about 8~yr, the MRI-active zone has reached its maximum extent, with a radius of 0.6~AU at the midplane and a vertical reach of up to z/r=$\pm0.08$ at 0.3~AU, which corresponds to 0.02~AU above and below the midplane ($\sim$ two scale heights). The time at which this stage is reached is denoted with $t_\mathrm{c}$. The disk's temperature map and surface density structure at that time are shown in the second column of Fig. \ref{fig:burststructure}. The yellow contour line indicates that the temperature within the ignited zone increases towards $T_\mathrm{S}$, which constitutes an equilibrium temperature. 

The solid blue line in Fig. \ref{fig:Tmid} shows the midplane temperature of \texttt{MREF} at $t=t_\mathrm{c}$. Only in the very inner parts at radii smaller than 0.1~AU does the midplane temperature increase slightly above $T_\mathrm{S}$ due to the large density of the accreted accumulated material and consequently significant viscous heat dissipation. The comparison with the profile of \texttt{M3} at $t=t_\mathrm{c}$ (solid red line) shows that in the more massive model, the ionisation front is able to travel farther and a larger part of the dead zone becomes MRI active. Hence, more mass is being accreted and the enhanced density of the accreting material in the inner disk allows the temperature to rise significantly above $T_\mathrm{S}$.
Notably, even when the TI is fully developed, there is still a thin dust arc between the ignited, viscously heated area and the inner gas disk. This arc is pushed close to the star, towards a radius of 0.055~AU in the case of \texttt{MREF}. It is manifested as a temperature sink close to the star, visible in the solid blue line in Fig. \ref{fig:Tmid}. The arc constitutes the inner dust rim, as explored in \cite{Flock2019}, which we resolve numerically using our description for the dust-to-gas ratio (Eq. \ref{eq:fd2g}). The rim travels closer to the star due to the large increase in density in the inner disk by the accretion process during the burst. In addition to $f_{\Delta \tau}$ and the resolution of the simulation, the thickness of the arc depends on the choice of the value of $\tau_*$ above which $f_\mathrm{D2G}$ is set to its maximum value $f_0$. In our models, this threshold value is set to 3.0. A larger value would lead to a thicker arc separating the inner gas disk from the ignited area since the increase in optical depth per unit length is limited by $f_{\Delta \tau}$ for larger radii compared to a smaller value. This limiting mechanism leads to a larger optically thin region beyond the dust sublimation front, which enables more efficient cooling and, therefore, increases the extent of the area where dust can exist. As soon as the $\tau_*$ threshold is reached, the dust-to-gas mass ratio increases quickly, creating an optically thick disk which is able to trap the viscously created heat and the temperature increases above $T_\mathrm{S}$ again. Although this arc is capable of blocking stellar irradiation, its existence and thickness do not change the thermal and dynamical evolution of the inner disk since the heating is dominated by viscous dissipation during the TI phase. We utilize this description of $f_\mathrm{D2G}$ in most of our simulations to be able to resolve the inner dust rim and stay consistent with the models of \cite{Flock2019}. However, we also created an alternative prescription, which is more suitable in the case of viscously dominated heating during the TI phase. We present and discuss the evolution of a model, including the new description in Sec. \ref{Sec:equilibrium_dust} and Appendix \ref{app:tau_thresh}.\\
After the ionisation front has reached its maximum radius at 0.6~AU, it travels back towards the star as a cooling front, gradually reinstating the dead zone due to the density being reduced by the accretion process and enabling efficient cooling to decrease the temperature below $T_\mathrm{MRI}$. The cooling front is defined analogously to the heating front, with the difference being that the cooling front propagates into the highly turbulent regions. The position of the heating/cooling front at the midplane is equivalent to $r_\mathrm{DZ}$ shown in panels b and b1 of Fig. \ref{fig:accr_rate_inset} during the TI development. After about 20 yr after the onset of the TI, the cooling front reaches a radius of $\sim$ 0.09~AU in the midplane on its way towards smaller radii (column 3 in Fig. \ref{fig:burststructure}). At that point in time, the TI develops yet again, reigniting the material in the dead zone slightly beyond the current position of the cooling front and causing the first reflare of the burst. The process of the reflare is equivalent to the one governing the first TI, the most significant difference being the smaller amount of material available for reignition after the previous accretion event. The maximum scale of the first reflare is reached after about 25 years after the first ignition, as is depicted in column 4 of Fig. \ref{fig:burststructure}. Multiple reflares follow with decreasing magnitude until a point where the TI can no longer develop due to inefficient production and trapping of viscous heat by the decreased density. The reflare process will be analysed in more detail in Sec. \ref{sec:S_curve_reflares}. Column 5 of Fig. \ref{fig:burststructure} shows the state of the disk at the time the TI has ended and the disk enters the quiescent phase again.

\subsection{Surface density evolution and pressure maxima} \label{sec:pressure_maxima}

\begin{figure*}[ht!]
    \centering
         \resizebox{\hsize}{!}{\includegraphics{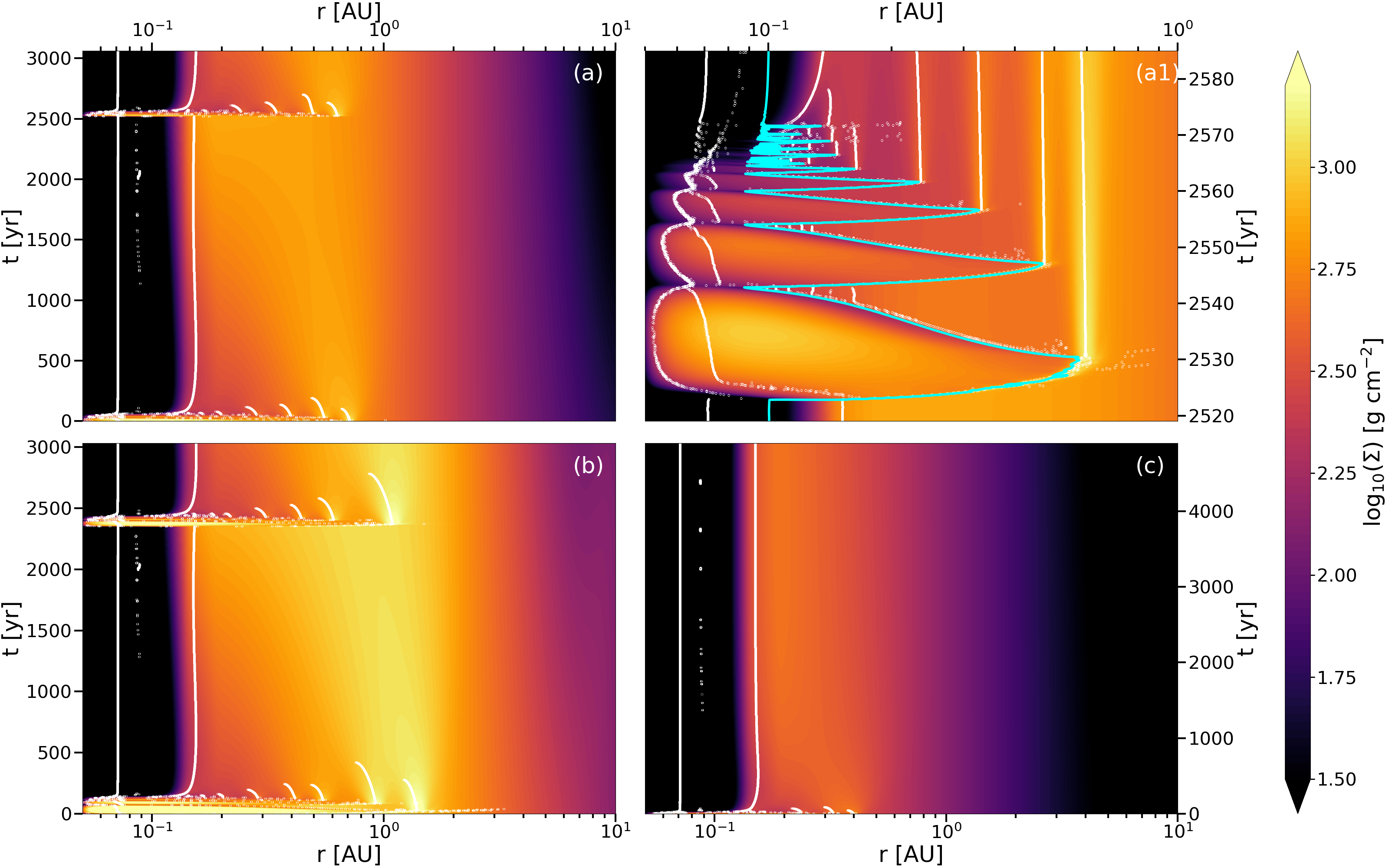}}
    \caption{
    Space-time diagrams of \texttt{MREF} (panel a), \texttt{M3} (panel b) and \texttt{M2} (panel c) with the colours corresponding to the surface density at the respective locations and times. While a TI cycle develops within a 3000 yr timeframe in the \texttt{MREF} and \texttt{M3}, the \texttt{M2} model does not show signs of TI even after about 5000 yr and the disk has reached a quasi-steady state.  The white contours indicate the positions of pressure maxima. Panel a1 shows the space-time diagram of the TI phase of the \texttt{MREF} model. The cyan contour line corresponds to the location of the inner edge of the dead zone at the midplane. 
    }
    \label{fig:r_t_tot}
\end{figure*}

The disk material is redistributed in the vicinity of the ionisation front according to the efficient angular momentum transport in the MRI active region. As a consequence, a sink in the surface density distribution develops behind the ionisation front while a corresponding spike travels ahead of the front. This behaviour is visible in the bottom panel of the second column in Fig. \ref{fig:burststructure}. As soon as the ionisation front has reached its largest radius and is reflected as a cooling front, the spike in surface density is left behind. Due to the decreasing magnitude of the consecutive reflares, every reflare places a density bump at the outer border of its respective MRI-active region. The result after the TI has ended is a `sawtoothed' surface density structure in the dead zone area ignited during the TI. This structure is shown in the bottom panel of column five in Fig. \ref{fig:burststructure}. Every spike in the surface density is associated with a corresponding maximum in the midplane gas pressure distribution. \\
Panel a of Fig. \ref{fig:r_t_tot} is a space-time diagram of the model \texttt{MREF} where the surface density is shown in colour. The white contour lines indicate the location of pressure maxima in the disk midplane. The stable contour during the quiescent phase at a radius of 0.15~AU is the pressure maximum equivalent to the pebble- and migration trap identified by \cite{Flock2019}. 

A zoom-in to the time frame of the TI developing after 2520 yr is provided in panel b of Fig. \ref{fig:r_t_tot}. In addition to the pressure maxima, the dead zone inner edge movement at the midplane is depicted as a cyan contour line. The pressure bumps are placed at the radii where the dead zone inner edge reaches its largest extent. During the burst, the stable pebble trap present in the quiescent phase is disrupted and develops again after the TI has ended. The midplane pressure maxima placed in the dead zone by the TI can be fitted by an analytical function of the form \citep[e.g., ][]{Taki2016, Lee2022, Lehmann2022},

\begin{equation}
    P_\mathrm{mid,fit}=P_0 \left (\frac{r}{1~\mathrm{AU}} \right )^a \left [ 1+b\, \mathrm{exp}\left (-\frac{(r-r_\mathrm{Pb})^2}{(w\,H(r_\mathrm{Pb}))^2} \right ) \right ] \; ,
\end{equation}

\noindent where $P_0 \left (\frac{r}{1~\mathrm{AU}} \right )^a$ describes the underlying power-law structure of the midplane pressure profile in the vicinity of $r_\mathrm{Pb}$ which is the radius of the pressure bump. $H(r_\mathrm{Pb})$ is the disk scale height at $r_\mathrm{Pb}$, calculated as $\Sigma(r_\mathrm{Pb})/\left (\sqrt{2\pi}\rho_\mathrm{mid}(r_\mathrm{Pb}) \right )$ with $\rho_\mathrm{mid}$ being the midplane density at the pressure bump location. $b$ and $w$ are dimensionless parameters, representing the amplitude and breadth of the density bump, respectively. For the pressure bump at $r=0.6$~AU, created by the first outward travel of the ionisation front, the fitted parameters $b$ and $w$ equate to 1.9 and 2.0, respectively. The density and pressure maxima decay rapidly due to viscous diffusion and disappear after 113 yr, which corresponds to about 240 orbits. In contrast, $b$ and $w$ for the second pressure bump at $r=0.47$~AU resulting from the first reflare yield 0.9 and 1.4, respectively, and the maximum decays after 150 yr or 460 orbits. \\
The space-time diagram for the model \texttt{M3} is shown in panel b of Fig. \ref{fig:r_t_tot}. In this case, the dead zone is more massive at $t=t_\mathrm{TI}$ and the ionisation front can travel further outwards. The TI starts to develop after 2350~yr and the first cycle deposits a pressure and density bump at a radius of 1.08~AU with the parameters $b$ and $w$, resulting in 2.0 and 1.7, respectively. The viscous accretion process drags the pressure bump towards 0.9~AU, where it disperses after 390~yr, which equates to about 400 orbits. \\
Panel c of Fig. \ref{fig:r_t_tot} depicts the space-time diagram for \texttt{M2}. Even after 5000 yr, the TI was not triggered and the pebble trap at 0.15~AU, slightly outside the location of the dead zone inner edge, persists undisturbed.

\subsection{S-curve and reflares}  \label{sec:S_curve_reflares}

\begin{figure}[t]
    \centering
         \resizebox{\hsize}{!}{\includegraphics{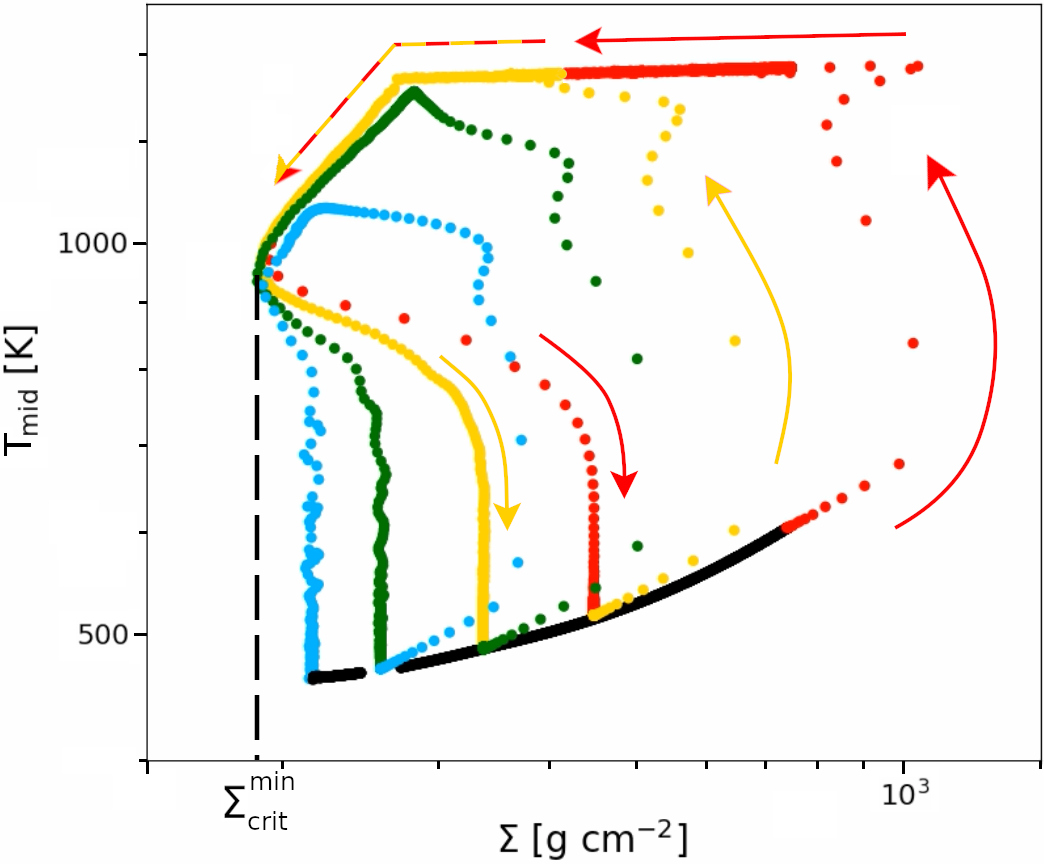}}
    \caption{
    S-curve of the \texttt{MREF} model at a radius of 0.2~AU. The black line corresponds to the evolutionary track during the quiescent phase, while the coloured dots are taken from snapshots during the outburst stage. The red data points were extracted during the first outburst and the yellow, green and blue dots make up the evolutionary tracks during the first, second and third reflare, respectively. The density of the dots along an evolutionary track indicates the velocity at which the disk evolves. The arrows are exemplary for the first two cycles and indicate the direction in which the model evolves. The surface density value at which the disk switches from the upper branch (where $\alpha=\alpha_\mathrm{MRI}$) to the lower branch (where $\alpha=\alpha_\mathrm{DZ}$) is the same for each cycle and is marked with $\Sigma_\mathrm{crit}^\mathrm{min}$.
    }
    \label{fig:S_curve}
\end{figure}

Every simulation computed for this work shows the occurrence of multiple reflares during the evolution of a TI-triggered episodic accretion event. The development of this phenomenon has been studied in the context of S-curves in the $\Sigma-T_\mathrm{eff}$ plane, which represent a visualisation of thermal stability (or instability) of the disk at a certain radius. In Fig. \ref{fig:S_curve}, we show such a S-curve for the model \texttt{MREF} at a radius of 0.2~AU. This location is situated in the dead zone, slightly outside the radius at which the TI is triggered. Instead of $T_\mathrm{eff}$, we show the midplane temperature $T_\mathrm{mid}$ since the midplane is where the TI elevates the temperature above $T_\mathrm{MRI}$ first during the expansion of the ionisation front. The lower stable branch (`low-state' of the disk), which corresponds to the MRI being inactive at this location (i.e. $\alpha=\alpha_\mathrm{DZ}$), consists of data points extracted during the quiescent state (black dots) and during the burst phase when the heating front has not reached 0.2~AU yet. As soon as the MRI is activated, the disk quickly switches to the upper branch of the S-curve where $\alpha=\alpha_\mathrm{MRI}$ (`high-state'). The maximum temperature of the upper branch is capped by the dust sublimation temperature, acting as an equilibrium state between heating and cooling. The surface density spike just before reaching the upper branch corresponds to the mass pile-up ahead of the ionisation front. As the mass is efficiently drained onto the star, the model moves along the upper branch towards smaller surface densities and smaller temperatures. After the kink in the upper branch, which is where the temperature drops below $T_\mathrm{S}$, the midplane temperature decreases further towards $T_\mathrm{MRI}=900$~K. While travelling along the upper branch, the tracks for the different cycles mostly overlap. $\Sigma_\mathrm{crit}^\mathrm{min}$ is the minimum value of the surface density at which the MRI can be sustained, which leads to the return of the model to the lower branch after passing $\Sigma_\mathrm{crit}^\mathrm{min}$. The S-curve shows that this critical value is the same for every cycle. After the last reflare (blue dots), the disk returns to the quiescent state and evolves along the lower branch. \\

\begin{figure}[t]
    \centering
         \resizebox{\hsize}{!}{\includegraphics{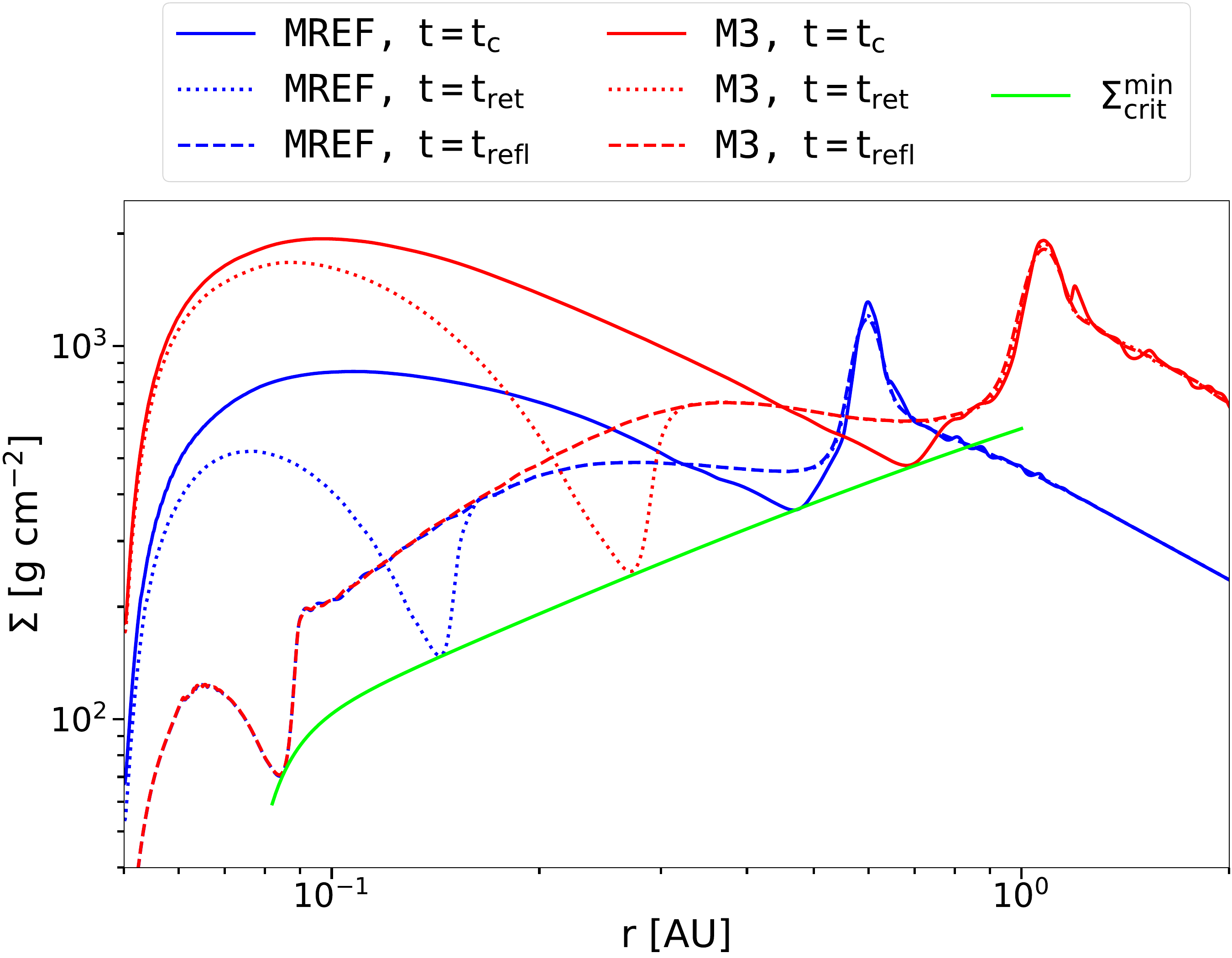}}
    \caption{
    Surface density profiles of the models $\tt{MREF}$ and $\tt{M3}$ for three points in time each. $t_\mathrm{c}$, $t_\mathrm{ret}$ and $t_\mathrm{refl}$ correspond to the points in time described in Fig. \ref{fig:accr_rate_inset}. The fit for the minimum values of the surface density valley travelling alongside the cooling front $\Sigma_\mathrm{crit}^\mathrm{min}$ at each radius is shown as a solid green line. 
    }
    \label{fig:reflare_fit}
\end{figure}

Fig. \ref{fig:reflare_fit} depicts the surface density profiles of \texttt{MREF} and \texttt{M3} at certain points in time during the burst phase. The times are also marked in panels a1 and b1 of Fig. \ref{fig:accr_rate_inset} for the example of \texttt{MREF}. $t_\mathrm{c}$ corresponds to the time at which the outwards moving ionisation front is reflected into an inwards moving cooling front. This occurs as soon as the minimum value in the surface density valley at the heating front reaches $\Sigma_\mathrm{crit}^\mathrm{min}$, which is shown as a green profile. Below $\Sigma_\mathrm{crit}^\mathrm{min}$, the MRI can no longer be sustained, $\alpha$ decreases, which diminishes the efficiency of viscous heating and the material cools down. This leads to the development of a cooling front with the density valley retreating towards the star. Snapshots of the surface density of \texttt{MREF} and \texttt{M3} during this time ($t=t_\mathrm{ret}$) are shown as dotted lines in Fig. \ref{fig:reflare_fit}. The minimum value of the retreating surface density valley is equal to the local value of $\Sigma_\mathrm{crit}^\mathrm{min}$. The fit for $\Sigma_\mathrm{crit}^\mathrm{min}$ shown in Fig. \ref{fig:reflare_fit} has the form,

\begin{multline}
    \Sigma_\mathrm{crit}^\mathrm{min}= 1678\, \left [\mathrm{g\, cm}^{-2} \right ]\, \left ( \mathrm{log}_{10}\left ( \frac{r}{1~\mathrm{AU}}+1 \right ) \right ) ^{0.856} \\
    -9\cdot 10^{-12}\, \left [ \mathrm{g\, cm}^{-2} \right ]\, \left ( \mathrm{log}_{10}\left ( \frac{r}{1~\mathrm{AU}}+1 \right ) \right ) ^{-8.582} \; ,
\end{multline}

\noindent where the power law component of this function outside of $\sim$0.12~AU is proportional to $r^{0.7}$.
The drop of the $\Sigma_\mathrm{crit}^\mathrm{min}$ profile at $r<0.1$~AU is related to the heating by irradiation: Close to the star, the importance of the irradiation heating relative to the heating by viscous dissipation increases, especially as the density in the inner disk decreases due to the accretion process. As a result, the surface density necessary to keep the disk MRI active becomes smaller. \\
In addition to the deviation of the $\Sigma_\mathrm{crit}^\mathrm{min}$ profile from a pure power-law close to the star, another effect of irradiation is the shift of the critical surface density for the ignition of a reflare, $\Sigma_\mathrm{crit}^\mathrm{max}$. This value is given by the position on the S-curve at which the disk leaves the lower branch. The corresponding point in time is denoted with $t_\mathrm{refl}$. In our models, all reflares are ignited at the same radius, so in order to visualise the different values of $\Sigma_\mathrm{crit}^\mathrm{max}$ for subsequent reflares, the S-curve can be evaluated at this radius. We emphasize that the reflares do not constitute a strict `reflection' of the cooling front into a heating front. Instead, the cooling front moves through towards the inner dust rim and the TI is triggered again (`reignited') in the cooled-down area, leading to another outburst cycle. On the other hand, the density wave travelling alongside the cooling front can be treated as being directly reflected. \\

\begin{figure}[t]
    \centering
         \resizebox{\hsize}{!}{\includegraphics{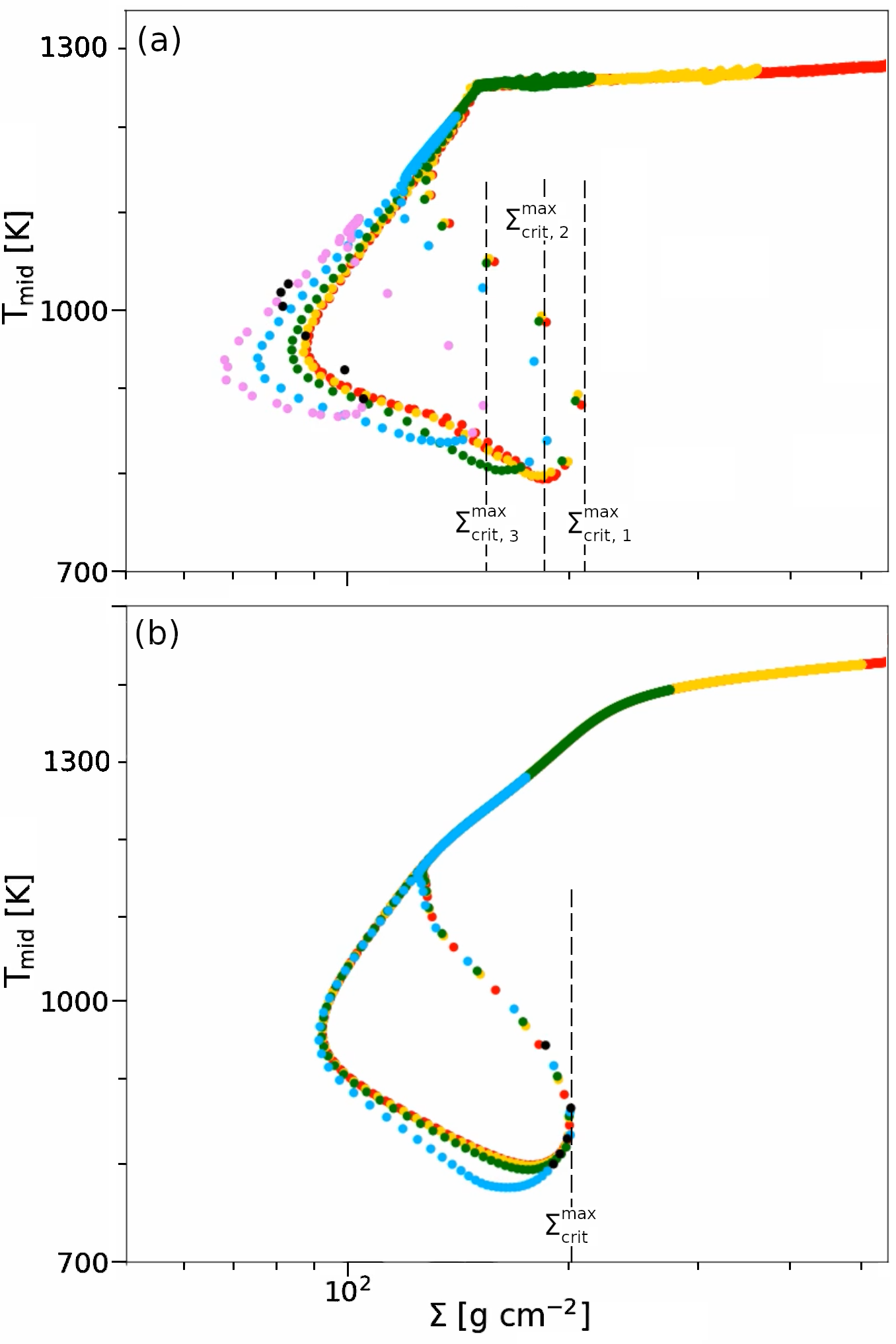}}
    \caption{
    S-curves during a TI-induced accretion event for the \texttt{MREF} model with (panel a) and without (panel b) the effect of irradiation heating, calculated at the respective radii at which the reflares are ignited. These radii are 0.091~AU and 0.087~AU for panel a and panel b, respectively. The data points with red, yellow, green, blue and pink colours were taken during the first, second, third, fourth and fifth reflare, respectively. With irradiation heating, the later reflares are ignited at smaller surface densities than the first three. The different critical surface density values are marked with $\Sigma_\mathrm{crit,\, 1}^\mathrm{max}$, $\Sigma_\mathrm{crit,\, 2}^\mathrm{max}$ and $\Sigma_\mathrm{crit,\, 3}^\mathrm{max}$. In the case without irradiation heating, all reflares are ignited at the same critical surface density value.
    }
    \label{fig:S_curve_no_irr}
\end{figure}

Analogously to Fig. \ref{fig:S_curve}, panel a of Fig. \ref{fig:S_curve_no_irr} shows the S-curve of the model \texttt{MREF}, but at a radius of 0.091~AU, which is where the reflares are reignited. Since this area of the disk never becomes fully MRI-inactive except for a very short time after the cooling front has passed through, the majority of the lower branch is missing in the depicted S-curve. Instead, the model switches back to the upper branch almost immediately after dropping down. The surface density value at which this occurs is not the same for every reflare, i.e. $\Sigma_\mathrm{crit}^\mathrm{max}$ is smaller for later reflares. While a smaller density leads to less effective viscous heating, it also reduces the radial optical depth from the star towards the reignition radius. Consequently, the $\tau=2/3$ surface is located further outwards. In combination with a steeper irradiation angle onto the $\tau=2/3$ surface at the reignition radius, this leads to more rigorous heating by irradiation, compensating for the decrease in the effectiveness of viscous heating. To test the effect of irradiation on the reflare behaviour, an additional simulation was conducted in which the heating by irradiation was switched off. Again, multiple reflares occur with the reignition radius being closer to the star by a small margin. The S-curve of this model at the reignition radius is shown in panel b of Fig. \ref{fig:S_curve_no_irr}. Without the effect of irradiation heating, the value of $\Sigma_\mathrm{crit}^\mathrm{max}$ is the same for every reflare.

\section{Discussion}\label{sec:discussion}
The simulations conducted in this study combine heating by viscous dissipation and irradiation with the evaluation of the dust-to-gas mass ratio and tracking of temperature-dependent levels of turbulence at every position in the computational domain. Our results show that for disk masses that are large enough, the inner disk is prone to thermal instability, which disrupts the vertical and radial temperature, density, and pressure structure. In this section, we analyse the implications of these disturbances and investigate the peculiarities of the TI phase in more detail. The analysis includes a discussion of the relevance of the pressure bumps emerging in our models, the variation of the dust content during the high-state of the disk and the potential reasons behind the occurrence of reflares. We also compare our models to previous studies, give an outlook on improvements of the $\alpha$-viscosity evolution description and point out limitations that our models are subject to. 
\subsection{Pressure bumps as possible sites for planetesimal formation} \label{sec:dis_pressure_bumps}
When the heating front travels through the inner disk towards larger radii, it sweeps over the area in which a large amount of mass has been accumulated during the quiescent phase. The strong increase in turbulent viscosity significantly enhances the transport of angular momentum in this high-density region. As a consequence, the heating front is accompanied by a density wave: The density behind the front is diminished while the material is compressed ahead of the front, creating a density bump \citep[`snowplough-effect', ][]{Lin1985}{}{}. When the heating front reaches its largest extent, this bump is left behind at that location while the cooling front retreats back towards the star. This maximum in density causes an analogous maximum in the gas pressure distribution, which, in turn, creates an area of super-Keplerian motion. The gas drag on dust particles then stops the inward flow of solid particles, creating a dust trap at the location of the pressure bump, similar to the dust trap created outside the orbit of a massive planet \citep[e.g.][]{Lambrechts2014}{}{}. In the cases described in this work, the pressure bumps evolve on the viscous timescale in the dead zone with a stress-to-pressure ratio of $\alpha_\mathrm{DZ}=10^{-3}$. \\
In order to estimate how much mass can potentially be trapped in the pressure bumps occurring in our models and how the density of dust may be altered at these locations as a consequence, we use the pebble flux predictor tool\footnote{\url{https://zenodo.org/records/4383154\#.YBE-TZzPwWo}} \citep[][]{Drazkowska2021}{}{} and calculate the mass accumulation in the strongest and most persistent pressure bump found in the model \texttt{M3}. The conditions in the disk around 1 AU, which is where the pressure bump is located, result in a pebble flux of $10^{-3} \mathrm{M}_\oplus ~\mathrm{yr}^{-1}$ and a Stokes number of $St=0.01$. As an initial pebble-to-gas mass ratio in the midplane, we assume 0.01. The pressure maximum persists for about 390 yr, which is enough time to potentially trap $ 0.39~\mathrm{M}_\oplus$ of pebbles. The dust sale height $H_\mathrm{d}$ can be calculated according to \citep[][]{Flock2021}{}{}, 

\begin{equation} \label{eq:dust_scaleheight}
    H_\mathrm{d}=\left ( \frac{H^2}{\frac{St ~ Sc}{\alpha_\mathrm{DZ}}+1} \right )^{1/2} \; ,
\end{equation}

\noindent where $H=c_\mathrm{s}/\Omega$ is the gas pressure scale height and $Sc$ is the Schmidt number, which is set to unity. We assume that the region in which the dust accumulates is a torus with a major radius of 1 AU and a minor radius of $H_\mathrm{d}$. Dividing the total accumulated mass after the pressure bump lifetime by the volume of the torus results in a mean dust density of $\rho_\mathrm{d, mean}=8.7\cdot 10^{-11}~\mathrm{g~cm}^{-3}$. Next, we impose a Gaussian distribution on the dust accumulating with a standard deviation of $H_\mathrm{d}/3$, so that the maximum dust density is located in the centre of the torus (i.e. at the midplane). This maximum value can be calculated by equating the volume under the Gaussian function to the volume of a cylinder with a base of radius $H_\mathrm{d}$ and a height of $\rho_\mathrm{d, mean}$,

\begin{equation}
    \rho_\mathrm{d, max}=\frac{9}{2}\rho_\mathrm{d, mean} \; .
\end{equation}

\noindent For the main pressure bump produced by the TI in the model \texttt{M3}, the maximum pebble density at the bump location yields $4\cdot 10^{-10}~\mathrm{g~cm}^{-3}$. The gas density in the area of the torus has values around $8.8\cdot 10^{-10}~\mathrm{g~cm}^{-3}$, which leads to a maximum solid-to-gas mass ratio of 0.45 at the midplane. Vertically integrating the Gaussian function for the pebble distribution at the pressure bump location results in a solid surface density of 100 g $\mathrm{cm}^{-2}$. Considering that the gas surface density at that location lies at 1900 g $\mathrm{cm}^{-2}$, this leads to a vertically integrated solid-to-gas mass ratio (also referred to as the metallicity $Z$) of 0.052. Following the empirical relation found by \cite{Yang2017} for $St<0.1$, 

\begin{equation}
    \mathrm{log}_{10}\,Z_\mathrm{crit}=0.1(\mathrm{log}_{10}\,St)^2+0.2(\mathrm{log}_{10}\,St)-1.76 \;,
\end{equation}

\noindent the critical solid-to-gas mass ratio above which significant spontaneous concentration of solids can occur results in a value of 0.017 for a Stokes number of 0.01. Since $Z>Z_\mathrm{crit}$ in the case of the analysed pressure bump occurring in our models, streaming instability may become efficient and can quickly raise the solid-to-gas mass ratio beyond unity 
\citep[e.g.,][]{Simon2016, Flock2021}{}{}. \\
The above calculation has been conducted under the assumption that the pressure bump stays at the same location and is capable of trapping all inward drifting solids throughout its entire lifetime. Therefore, the result should be seen as an upper limit. With the parameters of our model and the method described above, the pressure bumps created by the heating front during the TI may be capable of accumulating enough solids to induce streaming instability. Therefore, the TI mechanism analysed in our models can provide a possibility to form planetesimals over an extended radial range in the inner disk.
However, it's worth pointing out that the turbulent $\alpha$ parameter of $10^{-3}$ in the dead zone of our models may cause significant stirring around the midplane and inhibit the development of the streaming instability \citep[e.g., ][]{Li2022, Lesur2023}{}{}. On the other hand, the considerations above take the rather high value of $\alpha$ into account during the calculation of $H_\mathrm{d}$. Furthermore, the level of turbulence usually decreases when approaching the midplane \citep[e.g., ][]{Flock2017}{}{}, so $\alpha=10^{-3}$ might be an overestimation for the conditions near the midplane. Additionally, a higher Stokes number of 0.1 (i.e. assuming larger solid bodies) can significantly increase the amount of accumulated matter such that the solid-to-gas mass ratio at the end of the pressure bump's lifetime takes on values which favour the development of the streaming instability even more. A full evaluation of the conditions concerning streaming instability in the pressure bumps occurring in our models requires dedicated simulations, including coupled gas- and dust dynamics during the evolution of the pressure bump, as well as self-consistent treatment of local levels of turbulence. \\
Another possibility to form planetesimals in regions of enhanced dust concentration is the direct gravitational collapse of the dust cloud \citep[e.g., ][]{Johansen2006}{}{}. However, to fulfil the conditions for this formation process, the gas disk already has to be marginally gravitationally unstable \citep[e.g., ][]{Baehr2023}{}{}. In our models, the Toomre parameter in the disk regions around the density maxima placed by the TI evolution is of the order of $10^2$ even for the most massive disk model (\texttt{M3}). Following \cite{Gerbig2020}, a similar stability criterion can be formulated for the gravitational stability of a dust clump, which depends on the Toomre parameter, the metallicity $Z$, the efficiency of dust diffusion and the Stokes number. The conditions for gravitational instability are fulfilled when the stability parameter $Q_\mathrm{p}$ is smaller than unity. Applying this criterion to the strongest density bump in the model \texttt{M3} results in $Q_\mathrm{p}>10^2$. Therefore, we can assume that the dust concentrations formed in our models are gravitationally stable.

\subsection{An equilibrium dust density in the high-state region} \label{Sec:equilibrium_dust}
The description of the dust-to-gas mass ratio $f_\mathrm{D2G}$ given in Eq. \ref{eq:fd2g} is adequate for applications to irradiated inner disks with a resolved inner dust rim. However, a subtle difficulty arises when significant viscous heating during the TI phase is also considered. In the area of the disk in which $\tau_*>3$, a jump in $f_\mathrm{D2G}$ is manifested at $T_\mathrm{S}$, which does not take effect in a purely irradiated disk because temperatures in the vicinity of $T_\mathrm{S}$ are not expected beyond $\tau_*=3$. During a TI-induced accretion event, viscous heating is indeed capable of raising the temperature beyond $T_\mathrm{S}$ in the optically thick regions around the midplane. In these cases, when the temperature increases above $T_\mathrm{S}$, $f_\mathrm{D2G}$ switches from $f_0$ to a value dominated by $f_{\Delta \tau}$, which was only constructed for resolving the inner dust rim and should not have an influence on other disk regions. Since $f_{\Delta \tau}$ is small, the cooling efficiency abruptly increases, which leads to a stabilisation of the temperature at $T_\mathrm{S}$. In order to evaluate a more realistic dust density during the high-state of the disk, we updated the prescription for the dust-to-gas mass ratio,

\begin{equation} \label{eq:fd2g_alt}
    \begin{split}
    f_\mathrm{D2G}= & \left \{ f_{\Delta \tau} \frac{1}{8} \left [ 1- \mathrm{tanh}\left ( \left (\frac{T-T_\mathrm{S}}{150 \, \mathrm{K}} \right )^3 \right)\right] \left [ 1- \mathrm{tanh}(2/3-\tau_*)\right ] \right \}\\  
    & \cdot \left \{ 1+ \mathrm{tanh}\left (\frac{3.0-\tau_*}{0.4} \right ) \right \} \\
    & +\left \{ f_0\frac{1}{4}\left [ 1-\mathrm{tanh} \left ( \frac{T-T_\mathrm{S}}{20 \, \mathrm{K}}  \right) \right ] \right \} 
    \left \{ 1-\mathrm{tanh}\left (\frac{3.0-\tau_*}{0.4} \right ) \right \} \; .
    \end{split}
\end{equation}

This new function smoothly combines the evaluation of $f_\mathrm{D2G}$ in the region around the dust sublimation front ($\tau_*<3$) and the optically thick outer disk ($\tau_*>3$). There is now a smooth transition at $T_\mathrm{S}$ beyond $\tau_*=3$, where $f_\mathrm{D2G}$ is independent of $f_{\Delta \tau}$ and only depends on $f_0$ and the temperature. This allows $f_\mathrm{D2G}$ to smoothly and freely adopt a value that results in an equilibrium between heating and cooling in the high-state region. To analyse the effects of this new description on the adaptation of the dust density, we created two additional models \texttt{MREF}\textsuperscript{*} and \texttt{M3}\textsuperscript{*}, which are set up in the same manner as \texttt{MREF} and \texttt{M3} but include the function given in Eq. \ref{eq:fd2g_alt} for evaluating $f_\mathrm{D2G}$. In Appendix \ref{app:tau_thresh}, we illustrate the behaviour of this new description and show that its influence on the TI onset and evolution is minimal. \\

\begin{figure}[t!]
    \centering
         \resizebox{\hsize}{!}{\includegraphics{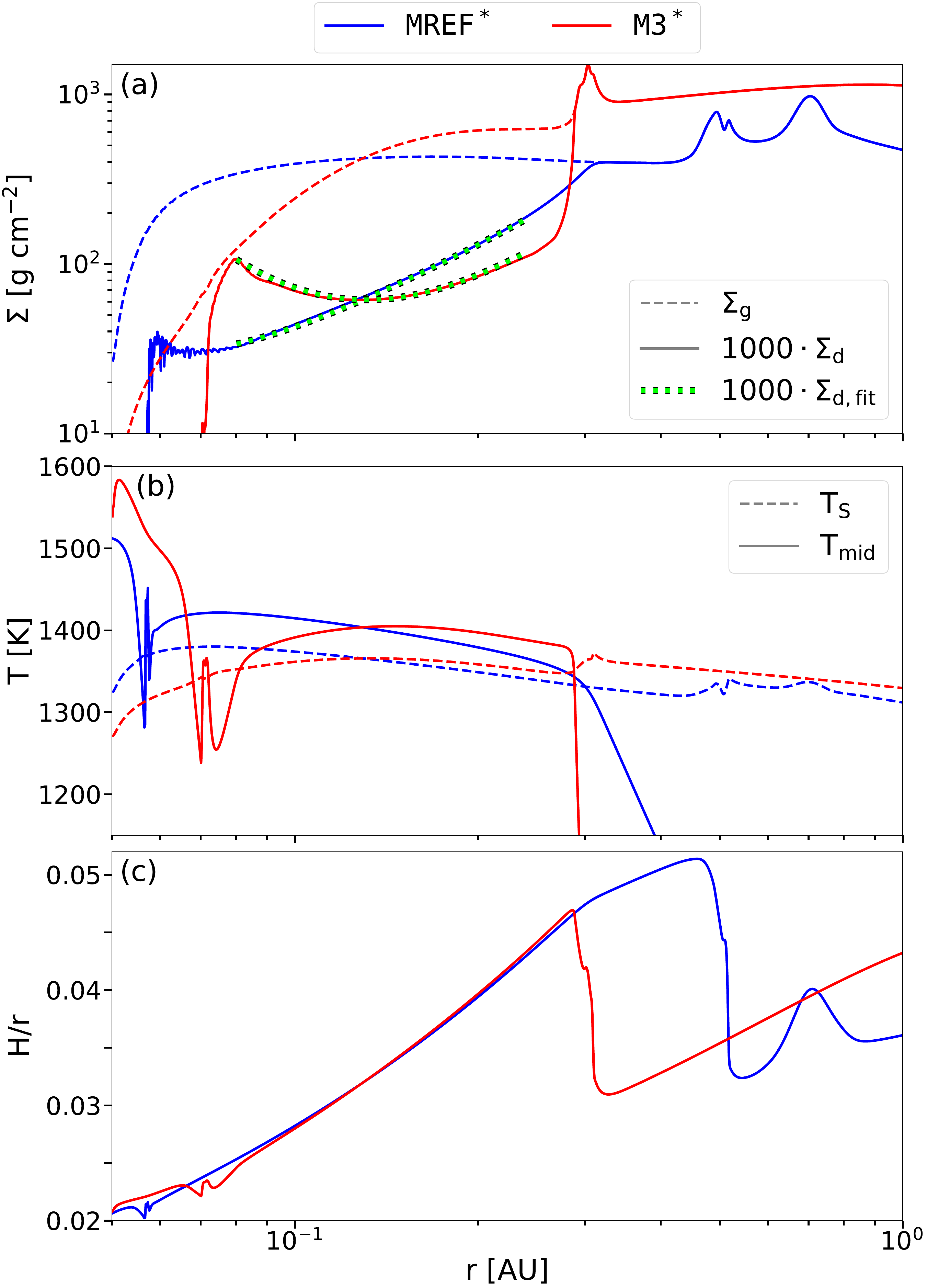}}
    \caption{
    Radial profiles of the surface density (panel a), the temperature (panel b) and the pressure scale height (panel c) within 1 AU for one representative point in time each during the TI development in the \texttt{MREF}\textsuperscript{*} and \texttt{M3}\textsuperscript{*} models. In panel a, dashed lines represent the gas surface density $\Sigma_\mathrm{g}$, solid lines show the dust surface density $\Sigma_\mathrm{d}$ and the dotted lines depict the respective fits to the dust surface density $\Sigma_\mathrm{d, fit}$ in the high-state region, evaluated with Eq. \ref{eq:Sigma_dust_fit}. $\Sigma_\mathrm{d}$ and $\Sigma_\mathrm{d, fit}$ are multiplied by 1000 to compensate for the baseline dust-to-gas mass ratio $f_0$ and make the values comparable to $\Sigma_\mathrm{g}$. The solid lines in panel b correspond to the midplane temperature, while the dashed lines represent the dust sublimation temperature.
    }
    \label{fig:newtau_dustfit}
\end{figure}

Fig. \ref{fig:newtau_dustfit} depicts radial profiles of various properties of \texttt{MREF}\textsuperscript{*} and \texttt{M3}\textsuperscript{*} during the evolution of an accretion event. The points in time that these profiles represent have been chosen such that the stage of the TI evolution is different: The profile of \texttt{MREF}\textsuperscript{*} has been taken at the time the first reflare has reached its largest extent, while the \texttt{M3}\textsuperscript{*} profile shows the stage of the disk during the first expansion of the heating front. Panel a shows the gas and dust surface densities. The radii at which the solid lines ($1000\cdot \Sigma_\mathrm{d}$, where $\Sigma_\mathrm{d}$ is the dust surface density) join the dashed lines (gas surface density $\Sigma_\mathrm{g}$) is the outer boundary of the region in which the temperature exceeds $T_\mathrm{S}$. The midplane- and dust sublimation temperatures for these two models are shown in panel b. The blue line indicates that $T_\mathrm{mid}$ has reached an equilibrium value in the highly viscous region when the heating front has reached its largest extent, and the inner disk has had enough time to adapt to the conditions on the upper branch of the S-curve. In the \texttt{M3}\textsuperscript{*} model snapshot, the accretion rate at small radii has not yet fully adapted to the high turbulence and the equilibrium temperature has only started to manifest itself. \\
We now aim to find a relation that connects the dust surface density in the fully ignited regions to other typical physical quantities. Since the dust surface densities seen in panel a of Fig. \ref{fig:newtau_dustfit} are a result of the equilibrium between heating and cooling in the respective regions, they have to depend on parameters which govern heating and cooling mechanisms in our disk model. In the high-state regions, heating is dominated by viscous dissipation, which mainly depends on the density, temperature and velocities (with the largest contribution coming from the angular velocity). The efficiency of radiative cooling is given by the optical depth, which again depends on the density and temperature around the ignited regions. Therefore, to first order, the dust surface density in the fully ignited, high-state equilibrium regions should be expressible in terms of the gas surface density $\Sigma_\mathrm{g}$ and the pressure scale height $H$. 
The profiles of the pressure scale height are depicted in panel c of Fig. \ref{fig:newtau_dustfit} for the two representative cases and show that the disk adapts itself so that the scale height at a given radius in the ignited disk regions in which $T>T_\mathrm{S}$ (i.e. in which an equilibrium temperature is reached) is the same for each model and every stage of the TI phase. A simple power-law fit to this profile of the scale height in these regions of the disk results in,

\begin{equation} \label{eq:H2r_fit}
    \left (\frac{H}{r} \right )_\mathrm{fit}=0.0283 \, \left (\frac{r}{0.1\, \mathrm{AU}} \right)^{0.482} \; .
\end{equation}

Based on this finding, the dependence of the dust surface density on $H$ can be transformed into a dependence on $r$. Fitting a power-law function to the dust surface densities in the high-state equilibrium regions results in the following relation,

\begin{equation} \label{eq:Sigma_dust_fit}
    \Sigma_\mathrm{d, fit}=34.8 \left (\frac{r}{0.1\, \mathrm{AU}} \right)^{1.75}\Sigma_\mathrm{g}^{-1.24} \, .
\end{equation}

The parameters given in the above equation have been evaluated based on a large number of snapshots at different stages during the TI evolution in the \texttt{MREF}\textsuperscript{*} and \texttt{M3}\textsuperscript{*} models. The green dotted lines in panel a of Fig. \ref{fig:newtau_dustfit} show the fit for one representative point in time for each model. \\
We emphasize that the relations given in Eqs. \ref{eq:H2r_fit} and \ref{eq:Sigma_dust_fit} are a result of the parameters chosen for the model setup described in this work. A different description of the dust- and gas opacities as well as a different choice of the viscosity parameters $\alpha_\mathrm{MRI}$ and $\alpha_\mathrm{DZ}$ can have a significant influence on the dependencies and numerical parameters in these expressions.

\subsection{Comparison to previous works}
Two-dimensional simulations of the protoplanetary disks in the $r-\theta$ plane during the TI stage have been carried out by \cite{Wunsch2005}, who also utilise a radiation-hydrodynamic model of a layered disk with a flux-limited diffusion approximation of the radiation transport and an MRI activation temperature of 1000 K. The main differences between their setup and our model are the opacity description, the absence of irradiation heating in their models and the size of their computational domain. Instead of separating the total density in the disk into a dust- and gas component via a prescription for the dust-to-gas mass ratio, they used the opacity relation given by \cite{Bell1993} for a gas-dust mixture. The results of their Model 5 most closely resemble the evolution of our models since they implemented the same $\alpha_\mathrm{MRI}/\alpha_\mathrm{DZ}$ ratio, which controls the rate of mass accumulation at the dead zone inner edge. The TI-induced accretion event occurring in their Model 5 is considerably less massive, entailing an amplification of the accretion rate by a factor of 6, compared to our cases with amplification factors of over 200. Interestingly, \cite{Wunsch2005} do not observe reflares in the same manner as they appear in our simulations. However, they derive a relation describing the critical surface density below which an MRI active disk cannot be sustained, which corresponds to $\Sigma_\mathrm{crit}^\mathrm{min}$ in the nomenclature of our work. They find that this critical value scales with $\sim r^{0.75}$, which is close to the scaling we derive for $\Sigma_\mathrm{crit}^\mathrm{min}$ in the region dominated by viscous heating ($\sim r^{0.7}$). \\
A similar relation has been derived by \cite{Nayakshin2024} for protoplanetary disks. Contrary to our model, the thermal instability occurring in their simulations is based on a strong increase in the gas opacity at temperatures larger than 2000 K \citep[as included in the opacity description given by ][]{Bell1993}{}{}, which is associated with the ionisation of hydrogen. In order to achieve such temperatures in the inner disk, they induce a continuous mass flux into the disk through the outer boundary at a rate of $>10^{-7}~\mathrm{M}_\odot \, \mathrm{yr^{-1}}$, which is not necessary for our models. They calculated S-curves for the vertical thermal stability of the disk structure at different radii and deduced a scaling for $\Sigma_\mathrm{crit}^\mathrm{min}$ and $\Sigma_\mathrm{crit}^\mathrm{max}$ with respect to the radius of $\sim r^{0.96}$. Considering that the TI mechanism has a different origin and the conditions in the disk, both in the low- and high-state, are different compared to our models, the discrepancy between their scaling relation and the one derived from our models is not surprising. However, reflares are occurring in some of their models. We will explore this phenomenon further in Sec. \ref{Sec:dis_reflares}. \\
The TI observed in the models of \cite{Nayakshin2024} is analogous to the classic disk instability model (DIM) for cataclysmic variables such as dwarf novae or X-ray transients \citep[e.g. review by ][]{Hameury2020}{}{}. In the context of protoplanetary disks, this kind of TI has been explored by \cite{Pavlyuchenkov2023}, who conclude that there are indeed multiple solutions for the disk's thermal structure under the conditions in which the gas opacity is strongly dependent on temperature. This results in the typical S-shape of the thermal equilibrium curves in the $\Sigma - T$ plane. If such a disk is evolved in time, including a viscous $\alpha$-prescrition, outbursts comparable to FU Ori events can occur. However, such conditions require surface densities and midplane temperatures that are not reached in our simulations. Nevertheless, the TI induced by the activation of the MRI with the descriptions used in our models results in a comparable limit cycle manifested as S-curves in the $\Sigma-T_\mathrm{mid}$ plane (see Fig. \ref{fig:S_curve}). In our models, the unstable branch of the S-curve, connecting the stable low-state and high-state equilibrium branches, is a consequence of the viscous $\alpha$ parameter (instead of the gas opacity) becoming dependent on temperature in the vicinity of $T_\mathrm{MRI}$ (according to Eq. \ref{eq_alpha}). This instability is also indicated in the models of \cite{Jankovic2021}. They report that in the regions of the inner disk where the MRI becomes quenched (corresponding to the inner edge of the dead zone), they find multiple solutions for an equilibrium of the vertical thermal structure. Similarly, \cite{Mohanty2018} find that their steady-state solutions might be viscously unstable due to a change in the vertically integrated stress-to-pressure ratio as a function of the mass flux. \\
Both versions of the instability were combined in a two-dimensional axisymmetric model by \cite{Zhu2009}. Since they aimed to explain and recreate observational signatures of FU Ori-type outbursts, the accretion rates, densities and temperatures are significantly higher compared to our models. Additionally, they do not consider irradiation by the central star or a fully MRI active disk at small radii. Instead, accretion in the innermost disk during quiescence is only driven by a mass flux in the ionized surface layers and the MRI is eventually activated beyond 1 AU by the accumulation of mass through an inward transport by gravitational instability, governed by a viscous $\alpha$ prescription. After the MRI becomes active and the accretion event is initiated, the temperatures increase so that the TI induced by hydrogen ionisation occurs as well. Although the objects considered by \cite{Zhu2009} are younger and more massive than the ones investigated in our simulations, and the burst is triggered by a different mechanism, the shape of the MRI active zone in the high-state during the outburst event generally agrees with our findings. \\
Another radiation hydrodynamic model of the inner regions of a protoplanetary disk in two-dimensional axisymmetric geometry was investigated by \cite{Flock2016}. The setup of their simulations is very similar to the one presented in this work. However, they considered Herbig Ae stars as the central objects, which inhibit luminosities of 56 $\mathrm{L}_\odot$. Consequently, the inner dust rim and the inner edge of the dead zone are shifted to much larger radii compared to our models, which consider a star with a luminosity of $\sim 2~\mathrm{L}_\odot$. While in our simulations, the stable pressure maximum at the dead zone inner edge during the quiescent state of the disk is situated at 0.15 AU, its location is shifted to 0.9 AU in the dynamic simulations of \cite{Flock2016}. At such distances, viscous heating is less effective and the required density to initiate the TI is much larger than the values they achieve. As a consequence, the disks do not become thermally unstable and remain in the low-state. \\
The development of pressure bumps as a result of accretion events induced by MRI activation in the dead zone has also been documented by \cite{Chambers2024}. They describe the evolution of fast- and slow-moving waves, which correspond to the movement of the heating/cooling front and the viscous evolution of the density maxima analysed in this work, respectively. The lifetime of the pressure bumps corresponding to the slow-moving waves is of the order of a few thousand years, which is longer than the existence of the pressure bumps occurring in our simulations after the TI. The reasons for this discrepancy are a larger value of $\alpha_\mathrm{DZ}$ implemented in our simulations, the smaller radius at which the bumps are situated initially and the absence of a pressure gradient force on the motion of the gas in the model of \cite{Chambers2024}. They speculate as well that the pressure maxima in their models may be capable of initiating streaming instability. We continue to build on this idea in Sec. \ref{sec:dis_pressure_bumps}.\\
\cite{Broz2021} investigated the formation of the architecture of the terrestrial planets in the solar system and found that models of convergent migration towards 1 AU provide a promising prospect. Interestingly, they indicate that the surface density structure required for such a migration mechanism should have a peak at 1 AU. Such a distribution looks remarkably similar to the disk structure of our models just after the TI phase (especially regarding the \texttt{M3} model, see panel d of Fig. \ref{fig:sigmacompare_mdots}). However, the surface density profile changes throughout the quiescent phase as the inner disk is replenished with material from larger radii and the peak is smoothed out (in the case of \texttt{M3}) or disappears (in the case of \texttt{MREF}, see panel c of Fig. \ref{fig:sigmacompare_mdots}). Only after a TI cycle a clear peak around 1 AU is reinstated. In order to combine the idea of convergent migration with a disk prone to outbursting events, dedicated simulations are required that include the migration of planetary bodies during and after the stage of the disk evolution in which outbursts occur.

\subsection{Possible origins of reflares} \label{Sec:dis_reflares}
Reflares are a well-known phenomenon in the context of disk instability models for cataclysmic variables \citep[][]{Lasota2001, Coleman2016, Hameury2020}{}{}. They are typically considered a weakness of the DIM since observations of rebrightenings after a main burst are rare for dwarf novae and X-ray transients. Reflares have also been reported in simulations of protoplanetary disks under certain conditions during the evolution of FU Ori-type outbursts \citep[][]{Nayakshin2024}{}{}. All models analysed in our work show the occurrence of reflares. \\
They are typically characterised as a `reflection' of the cooling front into a heating front. In our models, however, the disk region between the inner dust rim and the location at which the heating front starts moving outwards again also begins to cool down. The temperature then strongly increases again at a radius of 0.091 AU for the \texttt{MREF} and \texttt{M3} models, launching a new heating front in all directions, which can be described as a `reignition' of the TI rather than a reflection of the cooling front. As argued by \cite{Coleman2016}, the reason for this phenomenon could lie in the specific shape of the S-curve and the value of the viscous $\alpha$ parameter in the high-state of the disk. If the difference between $\Sigma_\mathrm{crit}^\mathrm{min}$ and $\Sigma_\mathrm{crit}^\mathrm{max}$ is too small, the surface density behind the inwards moving cooling front can eventually become larger than $\Sigma_\mathrm{crit}^\mathrm{max}$ and the TI starts anew. On the other hand, a very high value of $\alpha_\mathrm{MRI}$ enables more efficient draining of the ignited disk material on the star, which reduces the post-cooling-front density and could inhibit the occurrence of reflares. However, $\alpha_\mathrm{MRI}=0.1$, as used in our models, is already at the upper limit for MRI-active zones \citep[e.g., ][]{Flock2017}{}{}. A smaller value for $\alpha_\mathrm{DZ}$ could, in principle, shift $\Sigma_\mathrm{crit}^\mathrm{max}$ to larger values, therefore increasing the difference between the two critical surface density values. On the other hand, this would lead to a more massive outburst due to the large density necessary for triggering the TI and the heating front being able to travel to larger radii. Another factor for determining the shape of the S-curve and the critical surface density values is the function controlling the switch between $\alpha_\mathrm{MRI}$ and $\alpha_\mathrm{DZ}$ (Eq. \ref{eq_alpha}): A broader temperature range within which $\alpha$ is dependent on the temperature can contribute to the difference between $\Sigma_\mathrm{crit}^\mathrm{min}$ and $\Sigma_\mathrm{crit}^\mathrm{max}$ and can possibly have an influence on the reflare behaviour. \\
In addition to these factors, \cite{Nayakshin2024} find that reflares tend to disappear for smaller values of the critical temperature around which $\alpha$ changes between the low-state- and high-state values. A change in $T_\mathrm{MRI}$ does not influence the values of the critical surface densities significantly, but it has an impact on the shape and position of the unstable middle branch of the S-curve, which they deem vital for the emergence of reflares. In our models, we choose a relatively low value for $T_\mathrm{MRI}$ \citep[][]{Desch2015}{}{}{}. However, since the mechanism for triggering the TI in the setup of \cite{Nayakshin2024} is due to a strong dependence of the gas opacity on the temperature and the values for the critical temperature are of the order $10^4$\,K in contrast 900\,K used in our simulations, the comparison between their models and ours in the context of the occurrence of reflares might not be accurate.\\
Since the onset of the TI not only depends on the efficiency of viscous heating but also on the effectiveness of heat-trapping in the vicinity of the inner dead zone edge, it is worth considering the evolution of the opacity of the material near the midplane in the timeframe within which a reflare is ignited. Interestingly, a reflare is ignited at almost the same time at which the temperature drops below $T_\mathrm{s}$ at every location beyond the dust sublimation front. This suggests that a more effective heat-trapping near the midplane caused by a sudden increase in optical depth by condensation of dust may be responsible for the reignition of the TI. However, we conducted several tests in which the optical depth in the vicinity of the location of the reignition was kept at a low level artificially and did not observe significant changes in the reflaring behaviour. \\
Irradiation by the central star only marginally affects the emergence of reflares. As shown in Sec. \ref{sec:S_curve_reflares} and Fig. \ref{fig:S_curve_no_irr}, the location of the reignition is closer to the star and the two critical surface density values do not change for different reflare cycles if the heating by stellar irradiation is not considered. \\
The location of the computational domain's inner boundary is another relevant factor for investigating the occurrence of reflares. In our models, the innermost part of the disk is always on the upper branch of the S-curve, making it difficult for the cooling front to reach the inner boundary. The hot material in the inner MRI active disk creates a density- and corresponding pressure maximum while being accreted onto the star during the outburst cycle (see Figs. \ref{fig:reflare_fit} and panel a1 of Fig. \ref{fig:r_t_tot}). Even though the value of $\alpha_\mathrm{MRI}$ is chosen to be large, the accretion process is not effective enough to diminish this density- and pressure bump quickly enough. As the cooling front travels back towards the star, the pressure bump may still be too strong for the incoming density- and pressure waves to cross over. Therefore, the density wave may partly be reflected and could deposit additional material behind the cooling front, enough to trigger the MRI and launch a new heating front. If the inner boundary was chosen to be located at larger radii, more of the material making up the pressure bump may cross over the boundary and get accreted onto the star, making the reflection of the incoming density wave less likely. \\
In summary, reflares emerge due to the surface density behind the retreating cooling front crossing a critical surface density $\Sigma_\mathrm{crit}^\mathrm{max}$, which is set by the interplay between viscous heating near the disk midplane and the inefficiency of radiative cooling and consequent heat-trapping. Therefore, the questions of why the post-cooling-front surface density is high enough and why the value of $\Sigma_\mathrm{crit}^\mathrm{max}$ is low enough need to be considered when evaluating the origins of the reflaring behaviour. To answer these questions, investigations of the parameter space of possible $\alpha_\mathrm{DZ}$ and $\alpha_\mathrm{MRI}$ values, as well as variations and improvements of the descriptions for the opacities and dust-to-gas mass ratio, may be crucial. These considerations go beyond the scope of this work and will be part of future work.\\

\subsection{Thermal processing of solids}
\begin{figure}[t!]
    \centering
         \resizebox{\hsize}{!}{\includegraphics{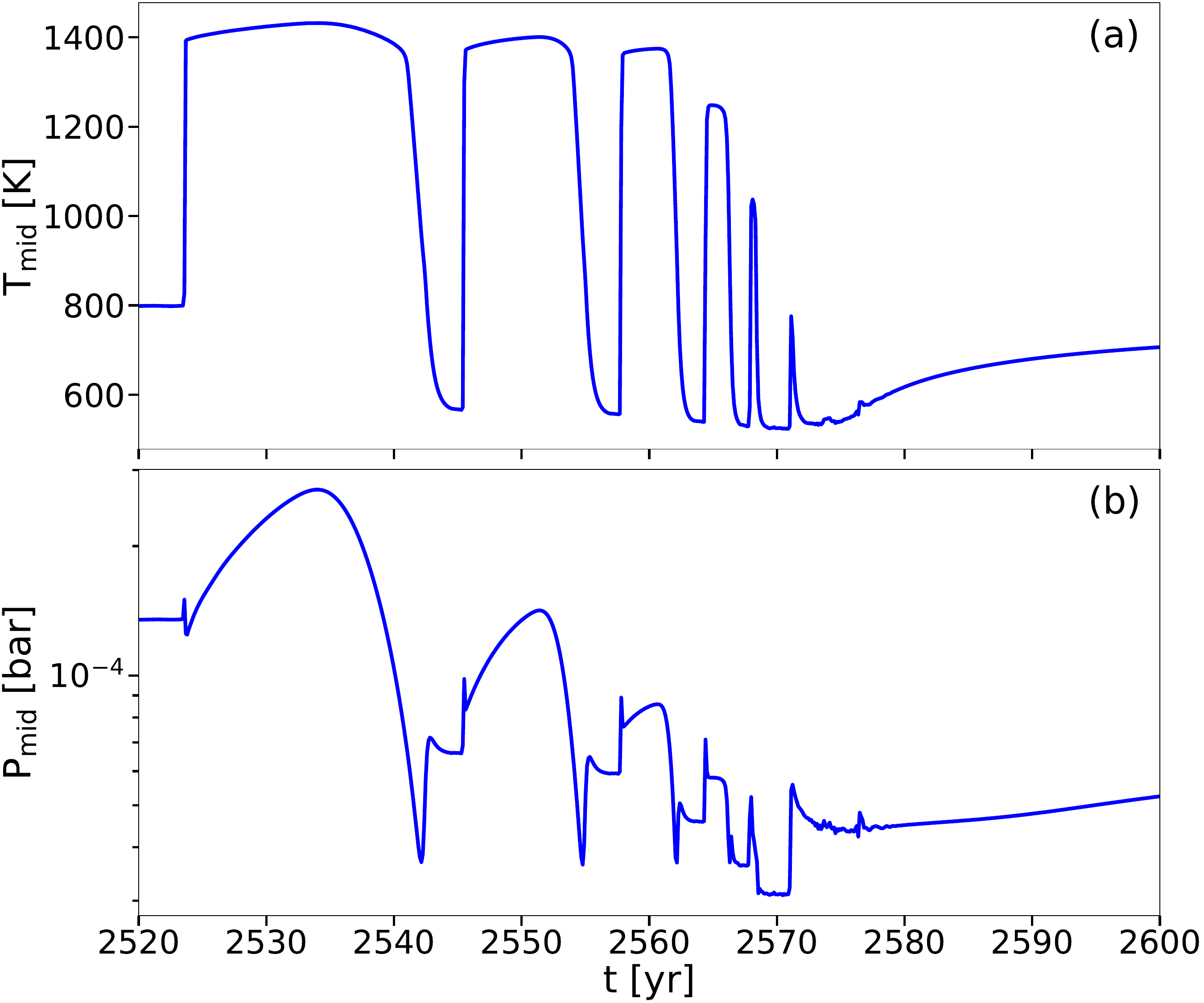}}
    \caption{
    Evolution of the midplane temperature (panel a) and midplane total gas pressure (panel b) at the location of the pebble trap (at 0.15 AU) during a TI-induced accretion event.}   
    \label{fig:bump_dust}
\end{figure}

The thermal history of a protoplanetary disk plays an important role in the nucleosynthetic composition of solid bodies in planetary systems \citep[e.g., ][]{Scott2007, Trinquier2009}{}{}. Hereby, chondrules and calcium-aluminium-rich inclusions (CAI) found in meteorites are of special interest since their occurrence and abundance give clues about the thermal conditions under which they are formed in the disk \citep[e.g. review by ][]{Scott2007}{}{}. The intricate composition of chondrules suggests that they have been subject to multiple rapid heating and cooling events during their formation, whereas condensation and subsequent coagulation of solids within $\sim$ 2 AU from the star seem to be necessary to explain the CAI abundance and the depletion of moderately volatile elements in chondritic meteorites \citep[][]{Cassen2001}{}{}. \cite{Connelly2012} argue that accretion processes might be capable of providing favourable conditions for chondrule formation in the inner disk. However, certain chondrule species only form late in the disk evolution (with disk ages of up to 3 Myr) when accretion rates are not thought to be large enough anymore to create the necessary high temperatures in the dusty disk regions. Dynamic crystallization experiments indicate that temperatures just below the chondrule melting temperature ($\sim$ 1600 K) are required to create specific textures found in chondrules \citep[e.g., ][]{Jones2018}{}{}. Our models show that episodic accretion processes can repeatedly elevate temperatures towards this regime. Fig. \ref{fig:bump_dust} shows the midplane temperature- and gas pressure modulation during the accretion event at 0.15 AU, which is where the pebble trap is situated during the quiescent phase. Following \cite{Flock2019}, we can assume that solid bodies accumulate at this location and may undergo the depicted temperature variations. 

Panel a indicates that the midplane temperature can change rapidly by several hundred Kelvin during the evolution of the TI. For instance, after the first burst, the temperature drops from $\sim$ 1400 K to below 600 K, after which it increases again during the first reflare towards 1400 K. The amplitude of the temperature fluctuations becomes smaller in later reflares. However, the required heating and cooling rates for chondrule formation are thought to be in the range of several 100 K $\mathrm{h}^{-1}$ \citep[e.g., ][]{Jones2018}{}{}, which is much larger than the ones resulting from our models. Here, the temperature increases by about 800 K over the course of one month while the cooling rates are even smaller. Furthermore, the temperatures in our models remain at levels close to the sublimation temperature for several years before cooling down, which is also in contrast to the required thermal processing of chondrules \citep[e.g., ][]{Libourel2018}{}{}. Therefore, our TI models may be unfit to provide the appropriate conditions for chondrule formation. \\
It is worth noting, however, that the explanation of chondrule structures and composition is more complex. \cite{Libourel2018} argue that the interaction of the solids with the surrounding gas component of the disk is essential and can allow for more moderate cooling rates. Furthermore, due to the highly dynamic process of the accretion event, it is unclear how the solid particles trapped near the inner dead zone edge are redistributed during and after the TI evolution. While a significant amount might be accreted onto the star, a part of the solid content may be transported to larger radii or higher disk layers by the expanding heating fronts and accompanying density waves. Therefore, the thermal history of individual particles cannot be inferred directly from the variations shown in Fig. \ref{fig:bump_dust}. Explicit inclusion of dust particles with varying Stokes numbers in our models could provide essential insights into the thermal and dynamic evolution of solids in the disk. \\
Additionally, panel b of Fig. \ref{fig:bump_dust} shows the variation of the midplane total gas pressure. For instance, \cite{Tsuchiyama1999} describe different evaporation regimes for forsterite in disks, depending on the total gas pressure and the temperature. Therefore, the self-consistent calculation of temperature and pressure in disk regions undergoing strong dynamic and thermal variations could help in the analysis of the mechanisms and consequences of the evaporation of solids.

\subsection{Model limitations}
Although our models incorporate multiple important aspects concerning the evolution of the inner rim during and between phases of thermal instability, specific limitations must be taken into account. \\
Our initial models represent hydrostatic structures of inner protoplanetary disks based on the setup of \cite{Flock2019}. They do not include radial and vertical velocity fields and are already thermally unstable due to the effect of viscous heating. As a consequence, the disk immediately enters the TI phase after initiating the time-integration of the hydrodynamic equations. Although the evolution of the initial thermal instability cycles closely resembles the ones occurring during the simulations, it is unclear to what extent they are influenced by our choice of the initial conditions. A steady-state initial model including velocity fields and with the disk located on the lower stable branch of the S-curve shortly before reaching $\Sigma_\mathrm{crit}^\mathrm{min}$ at the dead zone inner edge would certainly improve the quality of the beginning stages of the simulations. \\
In accordance with \cite{Flock2019}, the opacity description of our models consists of averaged values for both the gas- and dust opacities in order to properly resolve the inner dust rim. Using fitting functions for the opacities such as the ones given by \cite{Bell1993} or \cite{Zhu2009b}, or interpolating between pressure- and temperature-dependent values from tables \citep[as provided by e.g., ][]{Malygin2014}{}{} could have an influence on the magnitudes of the accretion events and the timescales of the quiescent and TI phases.\\
In our models, the fraction of dust in the disk is considered to determine the optical thickness of the disk material. However, we neglect the drift, settling, and growth of different dust species of different sizes. Studies have shown that TI-induced accretion events may have a significant impact on these processes \citep[e.g., ][]{Vorobyov2022}{}{}. Additionally, considering dust evolution and distribution may help in understanding the importance of the pressure bumps placed throughout the inner disk during the TI cycles \citep[e.g., ][]{Taki2016, Lehmann2022}{}{}. In a similar manner, the inclusion of the chemical evolution during and after outbursts can give valuable insights into the chemical composition of disks influenced by strong changes in the thermal structure \citep[e.g., ][]{Rab2017}{}{}.\\
In our models, the inner boundary of the computational domain is located at a fixed radius of 0.05 AU. However, \cite{Hartmann2016} argue that the description of disk accretion becomes inaccurate in the vicinity of the magnetic truncation radius, which is where the accretion becomes dominated by the stellar magnetic field. When equating the magnetic pressure from the stellar magnetic field to the ram pressure of the accreting material, the location of the truncation radius can be evaluated using the description given in \cite{Hartmann2016}. Implementing a typical stellar magnetic field strength of 1 kG \citep[e.g., ][]{Johns2007}{}{} and the stellar parameters of this work given in Tab. \ref{tab:MREF} yields radii for disk truncation between 0.02 AU and 0.1 AU depending on the accretion rate during our simulations. \cite{Steiner2021} utilised this description for their inner boundary and found that its variability in combination with torques exerted on the disk by the stellar magnetic field \citep[][]{Romanova2015, Romanova2018}{}{} can change the outbursting behaviour. \\
Our models do not consider the effect of accretion shock luminosity. Assuming that about 50\% of the gravitational energy of the accreted material is dissipated, the accretion luminosity in our models would increase from about 1\% of the stellar luminosity during the quiescent state to a factor of a few larger than the stellar luminosity at the peak of the outburst. Although the heating in the optically thick regions is dominated by viscous dissipation during the TI evolution, this amplification of the irradiation could possibly facilitate the advancement of the heating fronts and have an influence on the dust evolution \citep[][]{Vorobyov2022}{}{} and thermal processing of solids throughout the disk \citep[][]{Colmenares2024}{}{}. \\
Various magnetohydrodynamic effects can have a significant influence on the inner disk structure and evolution: Magnetocentrifugal winds can be a major driver of accretion by removing angular momentum from the disk close to the star \citep[][]{Bai2013, Lesur2023}{}{}. During episodic accretion events, these magnetically driven outflows can be strongly amplified \citep[][]{Konigl2011}{}{}. Additionally, the removal of mass by winds could possibly help stabilize the disk against TI by reducing the surface density.  Considerations of non-ideal magnetohydrodynamic effects in the context of the structure and evolution of the inner disk rim \citep[][]{Faure2014, Flock2017a}{}{} and the effectiveness of angular momentum transport via MRI \citep[][]{Bai2013, Simon2018, Rea2024}{}{} could certainly help improving our results as well. Moreover, \cite{Flock2013} have shown that a vertical profile of $\alpha$ dependent on the volumetric density replicates results from full radiation-magnetohydrodynamic simulations more accurately. Implementing such a vertical dependency could have an influence on the dynamics of the TI. \\
Furthermore, in our description of the stress-to-pressure ratio, an instantaneous response of $\alpha$ to a change in temperature is assumed. However, previous studies have indicated that the magnetic fields giving rise to the MRI only gradually build up and decay, leading to a significant delay between changes in the disk properties and the corresponding effect on the turbulent viscosity \citep[e.g.,][]{Hirose2009, Flock2017a, Ross2017, Held2022}{}{}. In order to account for this effect, the description of $\alpha$ needs to be modified with a time-dependent factor. While such a modification has already been implemented in the models of \cite{Zhu2010a} and \cite{Cleaver2023} in a simplified manner, the consideration of this effect in our simulations needs to be handled more carefully. The reasons for this claim are twofold: On the one hand, the TI evolution is mainly characterized by the outward expansion of the MRI active region. However, since the growth- and decay rate of the MRI is a function of the orbital timescale, the outwards propagation of the heating front may be heavily impeded by this effect. This can have significant consequences for the characteristics of the accretion event and the formation and location of pressure maxima. On the other hand, our models include a highly turbulent inner gas disk. As a consequence, a transition between MRI-active and -inactive regions is present throughout the entire simulation. During the quiescent phase, this transition region may show a more dynamic behaviour than indicated by the models of this work because small perturbations in the disk structure cannot be smoothed out quickly by an $\alpha$ instantly reacting to changes. Instead, they may even be amplified or lead to continuous oscillations of the density and temperature structure. Consequently, the dead zone inner edge may hardly ever be a stable structure of the kind indicated by our models during the quiescent phases.  Another possible result of these oscillations may be the variability of the accretion rate on timescales of the order of years and amplitudes of factors of a few. These interesting prospects are based on first tests of a viscosity description that considers the gradual saturation of the MRI in the context of our models and will be investigated in detail in a forthcoming paper.

\section{Conclusion} \label{sec:conclusion}
In this work, we utilize axisymmetric 2D radiation hydrodynamic simulations to investigate the vertical and radial structure and time-dependent evolution of the inner protoplanetary disk around a Class II T Tauri star. The models include an inner fully MRI active disk, a description for dust sublimation to resolve the inner dust rim, a transition towards a dead zone via a viscous $\alpha$ prescription, heating by stellar irradiation and viscous dissipation, and calculation radiation transport in vertical and radial direction. Starting from a hydrostatic initial state, we track the evolution of thermal instability triggered by MRI activation in the dead zone and analyse its effects on the disk properties. The main results of our study can be summarised as follows:
\begin{itemize}
    \item The inner disk is prone to thermal instability as long as accretion heating and heat-trapping near the midplane can elevate the temperature to values at which $\alpha$ becomes dependent on the temperature, which leads to runaway heating and consequent activation of the MRI. Initial accretion rates of $\geq 3.6 \cdot 10^{-9}~\mathrm{M}_\odot \; \mathrm{yr}^{-1}$ can be sufficient to accumulate enough material at the dead zone inner edge to trigger thermal instability.
    \item The evolution of the TI creates an oblate-torus-shaped MRI active region in the inner disk that can extend up to 1 AU in radial and 0.05 AU in vertical direction symmetrically around the midplane. Within this region, temperatures become high enough for dust to begin to sublimate. 
    \item The instability disrupts the otherwise stable planetesimal/migration trap at the dead zone inner edge. The corresponding pressure maximum is reinstated only after the TI phase has ended.
    \item Reflares are a robust outcome of our simulations. They occur due to the elevation of the surface density above a critical value behind the retreating cooling front as it approaches the inner dust sublimation region.
    \item The instability cycle and reflares can be tracked on S-curves of thermal stability from which critical surface density values for the activation and quenching of the TI can be extracted. 
    \item The burst and the consequent reflares place pressure maxima at the respective locations where the outwards travelling heating front stalls. These bumps may be capable of trapping solid particles and can remain in the disk long enough to initialise streaming instability.
    \item On the upper stable branch of the S-curve, the dust density adapts itself so that viscous heating and radiative cooling are in equilibrium. In the radial sections of the inner disk, where this is the case during the evolution of the TI, we find a simple relation between the gas- and dust-surface densities.
\end{itemize}
Resolving the two-dimensional structure of the inner disk while carefully considering vertical and radial transport of radiation vitally contributes to our understanding of the thermal and dynamic evolution of planet-forming regions undergoing thermal instability. Despite certain numerical and physical limitations, our models provide a robust foundation for potential future investigations, which include improved viscosity descriptions, consideration of dust evolution, implementation of magnetohydrodynamic effects and creation of synthetic observations. The elaboration of these prospects will further illuminate the effects of thermal instability on the formation and evolution of planets in protoplanetary disks. 

\begin{acknowledgements}
We would like to thank Zhaohuan Zhu and Doug Johnstone for fruitful discussions about this work's ideas. We thank the anonymous referee for their constructive and thoughtful comments. This research was supported by Deutsche Forschungsgemeinschaft (DFG) under Grant No. 517644750.
\end{acknowledgements}
\bibliographystyle{aa}
\bibliography{export}

\begin{appendix} \label{sec:appendix}

\section{Gas opacity} \label{sec:app_opac}
\begin{figure}[ht!]
    \centering
         \resizebox{\hsize}{!}{\includegraphics{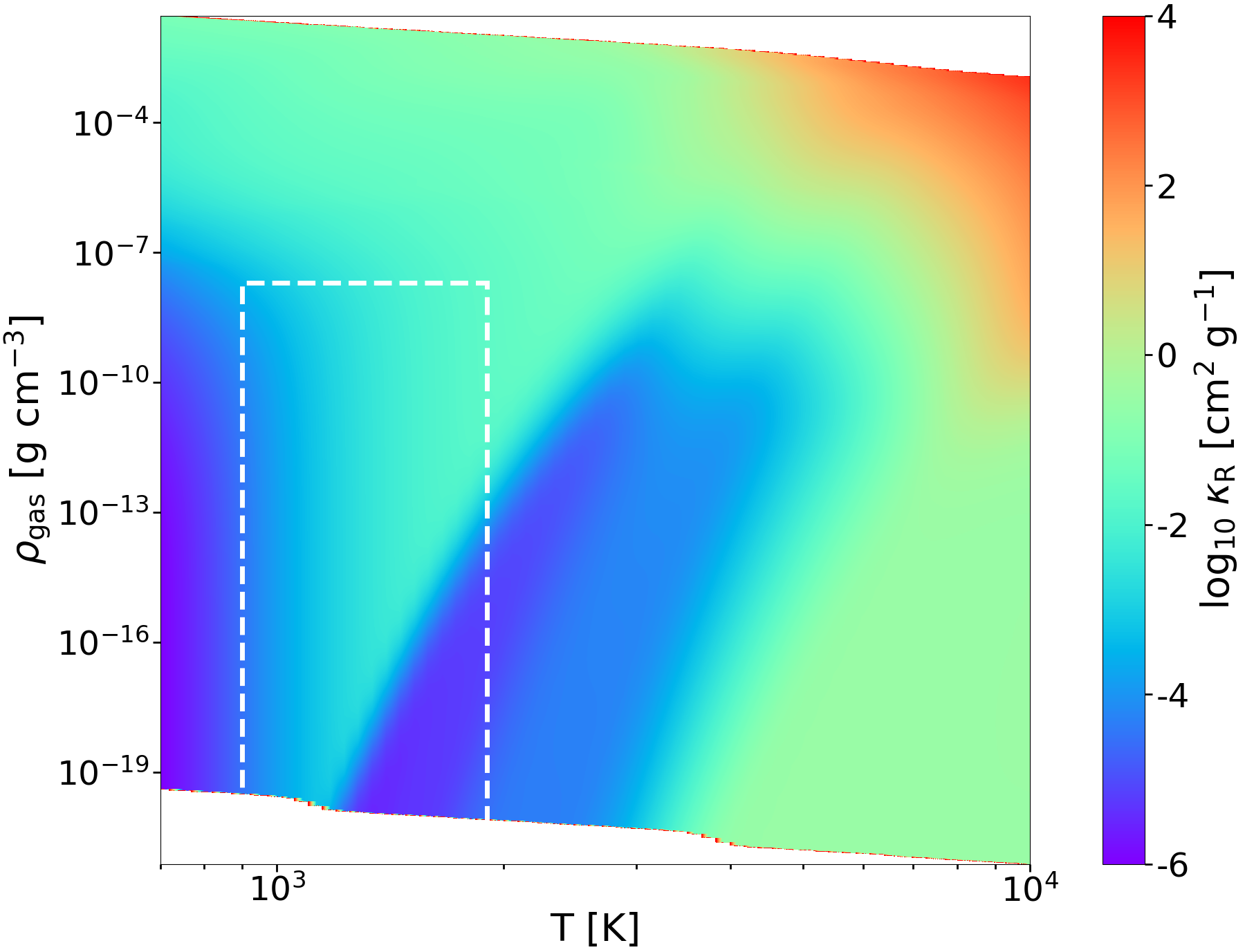}}
    \caption{
    Visualisation of the Rosseland mean gas opacity coefficients given by \cite{Malygin2014}. The white dashed rectangle indicates the density- and temperature regions for the gas from which the average value implemented in our simulations has been extracted.
    }
    \label{fig:opac_table}
\end{figure}

For our hydrodynamic simulations, we reevaluated the Rosseland mean gas opacity to account for more accurate values of gas density and temperature in the regions of the disk where gas dominates the opacity. In Fig. \ref{fig:opac_table}, we show the area in the $T-\rho_\mathrm{gas}$ plane within which the corresponding opacity values are considered. Temperatures below $\sim 800$ K are irrelevant for these considerations because, in this range, dust will always dominate the opacity. The average value in the outlined region in Fig. \ref{fig:opac_table} amounts to $\kappa_\mathrm{R}=10^{-3}\, \mathrm{cm}^2\, \mathrm{g}^{-1}$.

\vspace{-0.5cm}
\section{Updated computation of $f_\mathrm{D2G}$} \label{app:tau_thresh}
\begin{figure}[ht!]
    \centering
         \resizebox{\hsize}{!}{\includegraphics{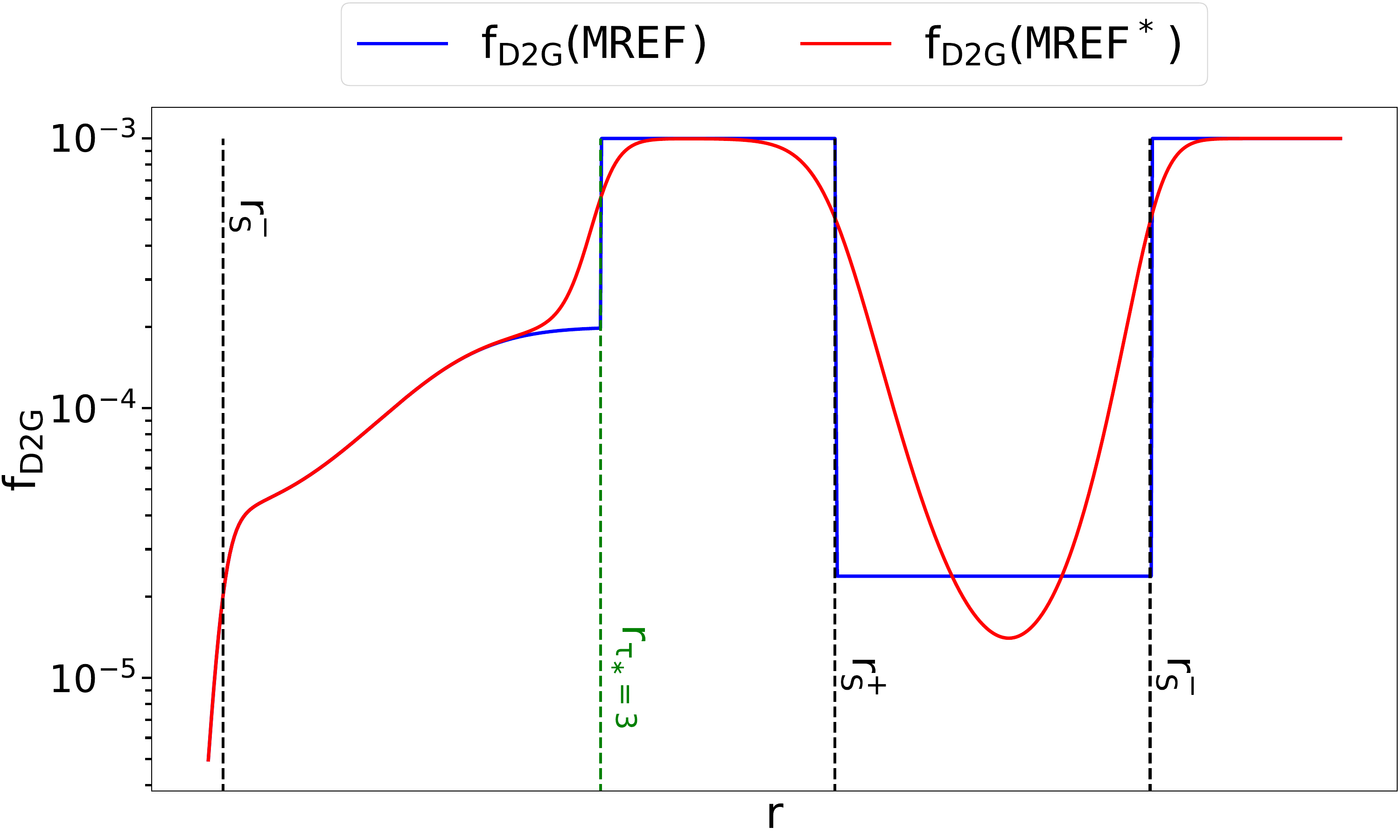}}
    \caption{
    Schematic visualisation of the functions describing the dust-to-gas mass ratio in the \texttt{MREF} and \texttt{MREF}\textsuperscript{*} models. For illustrative purposes, the density was held constant, representative profiles for the temperature and radial optical depth have been chosen and the models are shown as functions of a dimensionless radius. The vertical, black dashed lines mark locations where the temperature crosses the dust sublimation temperature. $r_\mathrm{S}^{-}$ and $r_\mathrm{S}^{+}$ denote crossings with a negative and positive temperature gradient, respectively. The green dashed line represents the radius at which the radial optical depth is equal to three.}
    
    \label{fig:ALTD2G_Scheme}
\end{figure}

\begin{figure}[h]
    \centering
         \resizebox{\hsize}{!}{\includegraphics{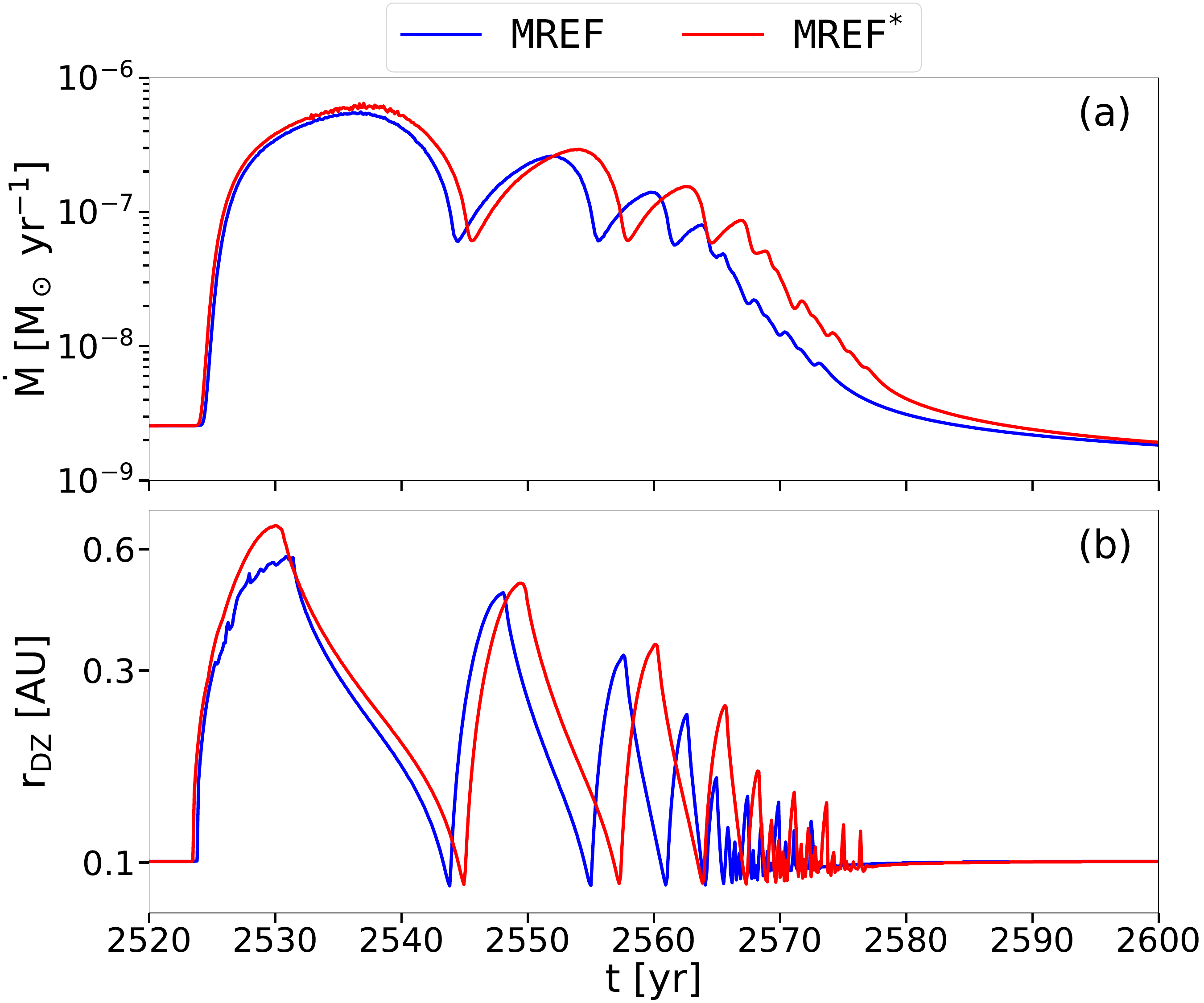}}
    \caption{
    Evolution of the accretion rate (panel a) and the position of the dead zone inner edge at the midplane (panel b) during TI-induced accretion events in the models $\texttt{MREF}$ and $\texttt{MREF}\textsuperscript{*}$.
    }
    \label{fig:ALTD2G}
\end{figure}

\begin{figure}[h]
    \centering
         \resizebox{\hsize}{!}{\includegraphics{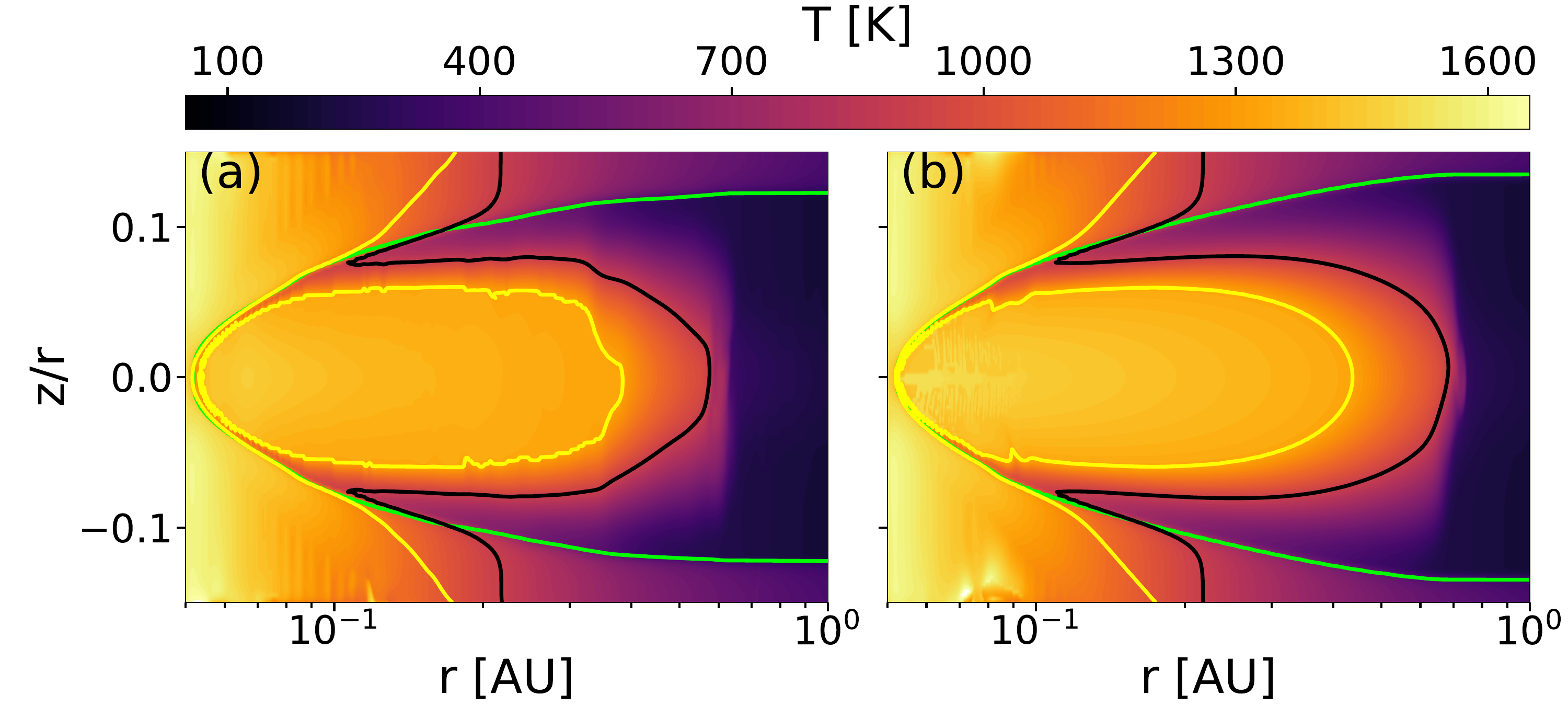}}
    \caption{
    Temperature map of the model \texttt{MREF} (panel a) and \texttt{MREF}\textsuperscript{*} (panel b) at the respective times at which the MRI active area has reached its largest extent and the cooling front starts to develop ($t=t_\mathrm{c}$). The contour lines are equivalent to the ones shown in Fig. \ref{fig:init_quiescent}.
    }
    \label{fig:ALTD2G_Tmap}
\end{figure}

In this section, we compare the new function for evaluating the dust-to-gas mass ratio, introduced in Sec. \ref{Sec:equilibrium_dust} and implemented in the model \texttt{MREF}\textsuperscript{*}, to the original one given by Eq. \ref{eq:fd2g}. Fig. \ref{fig:ALTD2G_Scheme} shows a schematic illustration of both functions, evaluated based on a constant density profile and representative temperature- and optical depth profiles. Before reaching the radius where $\tau_*=3$, the two functions do not differ significantly. Instead of a sharp jump to $f_0=10^{-3}$ at $\tau_*=3$ (which is largely exaggerated in Fig. \ref{fig:ALTD2G_Scheme}) in the original version, the new function produces a smooth transition. The largest difference reveals itself when the temperature crosses $ T_\mathrm{S}$ beyond $\tau_*=3$. While $f_\mathrm{D2G}$ instantaneously jumps to a value dominated by $f_{\Delta \tau}$ and locks the temperature at $T_\mathrm{S}$ in \texttt{MREF}, the new function lets $f_\mathrm{D2G}$ smoothly adapt as $T_\mathrm{S}$ is crossed. \\
Fig. \ref{fig:ALTD2G} and Fig. \ref{fig:ALTD2G_Tmap} show comparisons between \texttt{MREF} and \texttt{MREF}\textsuperscript{*}, conducted with the full simulation setup. While the quiescent stage is not influenced by the new function, the accretion event in \texttt{MREF}\textsuperscript{*} is marginally more massive and the MRI- and dust sublimation transitions are manifested in a smoother manner. Otherwise, the structure and evolution of the burst are not affected.

\end{appendix}

\end{document}